\documentclass[12pt,a4paper]{article}
\usepackage{jheppub_modified}

\allowdisplaybreaks[4]
\usepackage{amsmath,bm}
\usepackage{subfigure}

\title{
Black resonators and geons in AdS$_5$
}

\author[1,2]{Takaaki Ishii}
\author[3]{and Keiju Murata}
\affiliation[1]{Department of Physics, Kyoto University, Kyoto 606-8502, Japan}
\affiliation[2]{Institute for Theoretical Physics, Utrecht University, 3584 CC Utrecht, The Netherlands}
\affiliation[3]{Department of Physics, Osaka University, Toyonaka, Osaka 560-0043, Japan}
\emailAdd{ishiitk@gauge.scphys.kyoto-u.ac.jp}
\emailAdd{murata@het.phys.sci.osaka-u.ac.jp}

\abstract{%
We construct dynamical black hole solutions with a helical symmetry in AdS$_5$, 
called black resonators, as well as their horizonless limits, called geons.
We introduce a cohomogeneity-1 metric describing a class of black resonators and geons whose isometry group is $R\times SU(2)$.
This allows us to study them in a wide range of parameters.
We obtain the phase diagram for the black resonators, geons, and Myers-Perry-AdS$_5$, 
where the black resonators emerge from the onset of a superradiant instability of the Myers-Perry-AdS$_5$ with equal angular momenta
and are connected to the geons in the small horizon limit. 
The angular velocities of the black resonators always satisfy $\Omega>1$ in units of the AdS radius.
A black resonator is shown to have higher entropy than a Myers-Perry-AdS$_5$ black hole with the same asymptotic charges. 
This implies that the Myers-Perry-AdS$_5$ can dynamically evolve into the black resonator under the exact $SU(2)$-symmetry 
although its endpoint will be further unstable to $SU(2)$-violating perturbations.
}

\preprint{KUNS-2737, OU-HET-981}

\begin{document}
\maketitle

\section{Introduction}
\label{sec:intro}

Black holes and their gravitational dynamics in asymptotically anti de Sitter (AdS) spacetime have been widely studied recently.
Motivations for them partly come from the finding and development of the AdS/CFT duality \cite{Maldacena:1997re,Gubser:1998bc,Witten:1998qj}.
It has been revealed that the gravity in asymptotically AdS spacetime and higher dimensions has rich aspects to be explored.
(For instance, see \cite{Emparan:2008eg} for a review of higher dimensional black holes.)
To find black hole solutions is especially important for understanding the dynamics of the gravity in AdS.
Here, we focus on rotating black holes.
The higher dimensional generalization of the Kerr black hole solution, known as the Myers-Perry black hole~\cite{Myers:1986un}, 
was generalized to include a cosmological constant in five dimensions \cite{Hawking:1998kw} as well as in all dimensions \cite{Gibbons:2004uw,Gibbons:2004js}.
We will refer to a rotating black hole solution in higher dimensional AdS as Myers-Perry-AdS (MPAdS).
The thermodynamics of MPAdS was carefully discussed in \cite{Gibbons:2004ai} (see also \cite{Papadimitriou:2005ii}), 
and phase transitions for the rotating black hole solutions were studied in \cite{Carter:2005uw}.

Superradiance is a characteristic phenomenon of rotating black holes:
A wave scattered by a rotating black hole gains a larger amplitude than the injected wave under a certain condition. 
By the superradiance, energy and angular momenta can be extracted from the rotating black hole, and it makes 
the gravitational dynamics of the rotating black holes complicated.

In AdS, the superradiance leads to a dramatic scenario of gravitational dynamics known as the superradiant instability~\cite{Detweiler:1980uk,Cardoso:2004hs}.
Asymptotically AdS spacetime has a timelike boundary at conformal infinity, where waves are reflected.
Hence, for a rotating black hole in AdS, the wave amplified by the superradiance is injected back to the black hole.
By repetitions of the process, the initial perturbation grows exponentially.
In the Kerr-AdS$_4$ spacetime, a gravitational superradiant instability has been found in~\cite{Cardoso:2006wa}.
In the MPAdS spacetime, it has been found in \cite{Kunduri:2006qa,Murata:2008xr,Kodama:2009rq,Cardoso:2013pza} (see \cite{Brito:2015oca} for a review).

Once a superradiant instability occurs, new black hole solutions nonlinearly extending the instability are expected to appear.
At the onset of a superradiant instability, there is a normal mode.
In \cite{Kunduri:2006qa}, it was conjectured that there is a family of black hole solutions branching from the MPAdS as the nonlinear extension of the normal mode.
Such solutions have been numerically constructed in AdS$_4$ in \cite{Dias:2015rxy} and named {\it black resonators}. 
These black holes are time periodic and have only a single helical Killing vector. 

The first black resonators obtained in AdS$_4$ \cite{Dias:2015rxy} are found by solving three-dimensional partial differential equations (PDEs) 
and are cohomogeneity-3 (i.e.~the metric nontrivially depends on three coordinate variables).
As far as they were studied (see also \cite{Niehoff:2015oga}), 
the angular velocities of the black resonators always satisfy $\Omega>1$ in units of the AdS radius.
In~\cite{Green:2015kur}, it was shown that the AdS black holes with $\Omega>1$ are always unstable.
Therefore, the black resonators will have unstable modes.
However, their detailed analysis has not been done because of the difficulty due to few symmetries.
Hence, it would be desirable if there is a setup where the black resonators can be studied extensively.

In the zero-size limit of the black hole horizon, 
the superradiant frequencies reduce to normal modes in AdS.
Geons are then obtained as the nonlinear extension of the AdS normal modes.
They were constructed in asymptotically AdS spacetime both perturbatively and nonperturbatively~\cite{Dias:2011ss,Horowitz:2014hja,Martinon:2017uyo,Fodor:2017spc}.

In this paper, we construct a class of black resonators and geons which can be described by a cohomogeneity-1 metric in pure Einstein gravity in AdS$_5$.
We consider the five-dimensional MPAdS (MPAdS$_5$) with equal angular momenta as the ``background'' from which the black resonators appear.
This background spacetime has an $R\times U(2)$ isometry group.
We focus on the superradiant instability which preserves the $SU(2)$-symmetry of the MPAdS$_5$.
The resulting spacetime maintains $R\times SU(2)$ symmetries, and we find that it is described by a cohomogeneity-1 metric.
The Einstein equations hence reduce to ordinary differential equations (ODEs), and this allows us to study the black resonator and geon solutions in an extensive range of parameters.

The rest of this paper consists as follows.
In section~\ref{sec:mpads}, we start from reviewing the five-dimensional MPAdS$_5$ solutions and superradiant instability relevant to this paper.
In section~\ref{sec:cohomo1}, we introduce the metric ansatz for the cohomogeneity-1 black resonators and geons.
In section~\ref{sec:geon}, we construct the geons, first perturbatively and second fully numerically.
In section~\ref{sec:reso}, we obtain the black resonators and study their thermodynamics.
We conclude in section~\ref{sec:con} with a summary and discussions.

\section{Superradiant instability of Myers-Perry-AdS$_5$ with equal angular momenta}
\label{sec:mpads}

In this section, we review the five-dimensional Myers-Perry-AdS black hole (MPAdS$_5$) with equal angular momenta and its superradiant instability.
A black resonator is considered as the nonlinear extension of a rotating black hole's normal mode that exists at the onset of a superradiant instability.
We will use the equal-angular-momentum MPAdS$_5$ as the ``background'' to construct black resonators.
For that purpose, we first summarize the black hole solution and its perturbations.

\subsection{Myers-Perry-AdS$_5$ with equal angular momenta}

We consider the pure Einstein gravity in five dimensions with a negative cosmological constant,
\begin{equation}
S=\frac{1}{16\pi G_5}\int d^5x \sqrt{-g}\left[R+\frac{12}{L^2}\right]\ ,
\label{EHaction}
\end{equation}
where $G_5$ is the five-dimensional Newton's constant, $L$ is the AdS radius, and the cosmological constant is $\Lambda=-6/L^2$. 
Hereafter, we use units where $L=1$.
The Einstein equation is given by $G_{\mu\nu}-6 g_{\mu\nu}=0$.

While in general the MPAdS$_5$ has two angular momenta $J_1$ and $J_2$, in the case of equal angular momenta $J_1=J_2$, 
the metric of the MPAdS$_5$ can be given in a cohomogeneity-1 form \cite{Gibbons:2004uw},\footnote{%
These coordinates are brought to the Boyer-Lindquist coordinates in \cite{Gibbons:2004uw,Gibbons:2004js} by 
$r^2 \to \frac{r^2+a^2}{1-a^2}, \, \theta \to 2 \theta, \, \phi \to \varphi_1 - \varphi_2, \, \psi \to \varphi_1 + \varphi_2$, and $\mu \to \frac{M}{(1-a^2)^3}$ 
where $\psi$ will be introduced later.
}
\begin{multline}
 ds^2=-(1+r^2)f(r)d\tau^2 + \frac{dr^2}{(1+r^2)g(r)}\\+\frac{r^2}{4} \left[
\sigma_1^2 + \sigma_2^2 + \beta(r)(\sigma_3+2 h(r)d\tau)^2
\right]\ ,
\label{KerrAdS}
\end{multline}
where the metric components are given by
\begin{equation}
\begin{split}
&g(r)=1-\frac{2\mu (1-a^2)}{r^2(1+r^2)} +\frac{2a^2\mu}{r^4(1+r^2)}\ ,\quad
\beta(r)=1+\frac{2 a^2\mu}{r^4}\ ,\\
&h(r)=\Omega-\frac{2\mu a}{r^4+2 a^2\mu}\ ,\quad 
f(r)=\frac{g(r)}{\beta(r)}\ .
\end{split}
\label{KerrFunctions}
\end{equation}
The outer horizon of the black hole is located at $r = r_h$ that is the largest real root of $g(r_h)=0$. 
The constant $\Omega$ may be shifted by a coordinate transformation. 
For later convenience, we choose $\Omega$ as follows so that $h(r_h)=0$ is satisfied: 
\begin{equation}
\Omega = \frac{2\mu a}{r_h^4+2 a^2\mu}\ .
\label{OmegaMPAdS}
\end{equation}
With this choice, we have $g_{\tau\tau}|_{r=r_h}=0$.
This implies that $\partial_\tau$ is the null generator of the horizon.
In Eq.\eqref{KerrAdS}, we have introduced the one-forms $\sigma_a\,(a=1,2,3)$ on $S^3$ as
\begin{equation}
\begin{split}
&\sigma_1 = -\sin\chi d\theta + \cos\chi\sin\theta d\phi\ ,\\
&\sigma_2 = \cos\chi d\theta + \sin\chi\sin\theta d\phi\ ,\\
&\sigma_3 = d\chi + \cos\theta d\phi  \ . 
\end{split}
\label{invf}
\end{equation}
We also find it convenient to define
\begin{equation}
\sigma_\pm = \frac{1}{2}(\sigma_1 \mp i \sigma_2)=\frac{1}{2}e^{\mp i\chi}(\mp i d\theta + \sin\theta d\phi)\ .
\end{equation}
The one-forms satisfy the Maurer-Cartan equation $d\sigma_a - 1/2 \epsilon_{abc} \sigma_b \wedge \sigma_c = 0$.
The angular coordinates $\phi$ and $\chi$ are defined on a twisted torus:
The coordinate ranges are $0\leq \theta \leq \pi $, $0\leq \phi <2\pi$, and $0\leq \chi <4\pi$ with the periodicity
\begin{equation}
(\theta,\phi,\chi) \sim (\theta,\phi+2\pi,\chi+2\pi) \sim (\theta,\phi,\chi+4\pi) \ .
\label{phipsi_id}
\end{equation}
There are $SU(2)$ generators $\xi_i$ $(i=x,y,z)$ defined by
\begin{equation}
\begin{split}
 &\xi_x = \cos\phi\partial_\theta +
 \frac{\sin\phi}{\sin\theta}\partial_\chi -
 \cot\theta\sin\phi\partial_\phi\ ,\\
 &\xi_y = -\sin\phi\partial_\theta +
 \frac{\cos\phi}{\sin\theta}\partial_\chi -
 \cot\theta\cos\phi\partial_\phi\ ,\\
 &\xi_z = \partial_\phi\ ,
\end{split}
\label{su2_xi}
\end{equation}
which satisfy $[\xi_i,\xi_j]=\epsilon_{ijk}\xi_k$. 
The one-forms $\sigma_a$ satisfy $\mathcal{L}_{\xi_i}\sigma_a=0$ where $\mathcal{L}_{\xi_i}$ is the Lie derivative along the curve generated by the vector field $\xi_i$. 
Thus $\sigma_a$ are invariant 1-forms of this $SU(2)$. 
We have $\sigma_1^2+\sigma_2^2=d\theta^2+\sin^2\theta d\phi$, which gives the metric of a round $S^2$.
Then, in Eq.(\ref{KerrAdS}), an $r=$constant surface is given by a squashed $S^3$ written as 
a $S^1$ bundle over $S^2$.

There is another non-trivial Killing vector $\partial_\chi$. 
This vector field generates a rotation of $\sigma_1$ and $\sigma_2$,
\begin{equation}
\mathcal{L}_{i\partial_\chi}\sigma_\pm=\pm \sigma_\pm\ ,\quad \mathcal{L}_{i\partial_\chi}\sigma_3= 0\ ,
\end{equation}
i.e., $\sigma_\pm$ and $\sigma_3$ have $U(1)$-charges $\pm 1$ and $0$, respectively.
The $S^2$ part of the metric, $\sigma_1^2+\sigma_2^2=4\sigma_+ \sigma_-$, 
is invariant under $\partial_\chi$. 
Had there be no rotations, the round $S^3$ formed by $\sigma_{1,2,3}$ has had another $SU(2)$ symmetry which contains this $\partial_\chi$, 
but in \eqref{KerrAdS} it is broken down to $U(1)$ due to the angular momentum.
In summary, the isometry group of the geometry~(\ref{KerrAdS}) is given by 
$R_\tau \times SU(2) \times U(1) \simeq R_\tau \times U(2)$, 
where $R_\tau$ is the shift symmetry along the $\tau$-direction.

The thermodynamics of MPAdS was discussed in \cite{Gibbons:2004ai,Papadimitriou:2005ii}.
For the MPAdS$_5$ with equal angular momenta, 
the mass $E$, angular momentum $J$, entropy $S$, and temperature $T$ are given by
\begin{equation}
\begin{split}
&E=\frac{\pi}{4 G_5}\mu(a^2+3)\ ,\quad J=\frac{\pi}{G_5}\mu a\ ,\\
&S=\frac{\pi^2r_h^3}{2 G_5}\sqrt{1+\frac{2\mu a^2}{r_h^4}}\ ,\quad 
T=\frac{2(1-a^2)r_h^4+(1-4a^2)r_h^2-2a^2}{2\pi r_h^2\sqrt{(1-a^2) r_h^2-a^2}}\ .
\end{split}
\label{mpads_thermo}
\end{equation}
The angular velocity $\Omega$ is given by Eq.(\ref{OmegaMPAdS}).

\subsection{Rotating and non-rotating frames at infinity}
\label{sec:frames}

The solution given above actually corresponds to the {\em rotating frame} at infinity.
The asymptotic form of the MPAdS$_5$ metric at the AdS boundary $r \to \infty$ becomes
\begin{equation}
 ds^2\simeq -(1+r^2)d\tau^2 + \frac{dr^2}{1+r^2}+\frac{r^2}{4} \left[
\sigma_1^2 + \sigma_2^2 + (\sigma_3+2 \Omega d\tau)^2 \right]\ .
\label{MPAdSasym}
\end{equation}
Meanwhile, we took $h(r_h) = 0$ on the horizon, and we deduce for the coordinate $\tau$ that $\partial_\tau$ becomes the null generator of the horizon.

We can move to the {\em non-rotating frame} at infinity by defining $(t,\psi)$ as 
\begin{equation}
dt=d\tau\ ,\quad d\psi=d\chi+2 \Omega d\tau\ .
\label{psishift}
\end{equation}
In these coordinates, the asymptotic form of the metric becomes
\begin{equation}
 ds^2\simeq -(1+r^2)dt^2 + \frac{dr^2}{1+r^2}+\frac{r^2}{4} \left[
\bar{\sigma}_1^2 + \bar{\sigma}_2^2 + \bar{\sigma}_3^2 \right]\ ,
\label{metricasym2}
\end{equation}
where we introduced the invariant one-forms for the non-rotating frame
$\bar{\sigma}_a$ by replacing $\chi$ with $\psi$ in Eq.(\ref{invf}):
$\bar{\sigma}_1 = -\sin\psi d\theta + \cos\psi\sin\theta d\phi$, and so on.
The explicit relation between $\sigma_a$ and $\bar{\sigma}_a$ is 
\begin{equation}
\sigma_\pm = e^{\pm 2i\Omega t} \bar{\sigma}_\pm\ ,\quad 
\sigma_3=\bar{\sigma}_3-2\Omega d t\  .
\end{equation}
However, the combination $\sigma_1^2 + \sigma_2^2 = \bar\sigma_1^2 + \bar\sigma_2^2$ is invariant under the rotation.
The null generator of the horizon is given by the linear combination of $\partial_t$ and $\partial_\psi$ as\footnote{%
We define the angular velocity $\Omega$ with respect to $\psi/2 \in [0,2\pi)$ instead of $\psi$ itself. 
This $\Omega$ matches the definition in other literature~\cite{Gibbons:2004ai,Kunduri:2006qa,Murata:2008xr}.
}
\begin{equation}
 \frac{\partial}{\partial \tau} =  \frac{\partial}{\partial t}+\Omega\, \frac{\partial}{\partial (\psi/2)}\ .
\label{hericalKillingMPAdS}
\end{equation}
The trajectory of the horizon generator $\partial_\tau$ is rotating with the angular velocity $\Omega$ with respect to the non-rotating frame. 
This shows that Eq.(\ref{OmegaMPAdS}) gives the angular velocity of the event horizon.
In the rotating frame, we can view $\Omega$ as the boundary source of the rotation in the context of the AdS/CFT duality.
In fact, $\Omega$ is identified as the rotation chemical potential.

\subsection{Superradiant instability}
\label{SRMPAdS}

Let us consider a linear perturbation of the MPAdS$_5$ with equal angular momenta.
The gravitational superradiant instability of the MPAdS$_5$ has been found first in~\cite{Murata:2008xr}, and 
its detailed analysis also has been done in~\cite{Cardoso:2013pza}.
Here we focus on a $\tau$-independent perturbation of the form\footnote{
We would like to point out that this perturbation actually corresponds to charged tensor harmonics on $CP^1\simeq S^2$.
It has been widely believed that there are no charged tensor harmonics on $CP^1$ 
because the transverse traceless conditions and eigenvalue equation appear to 
overconstrain the charged tensor harmonics. 
In appendix~\ref{chargedTV}, we show that those conditions are not independent and then explicitly construct the charged tensor harmonics on $CP^1$.
We are grateful to Prof.~Harvey Reall for private communication.
}
\begin{equation}
\delta g_{\mu\nu}dx^\mu dx^\nu= \frac{r^2}{4} \delta \alpha(r) (\sigma_1^2-\sigma_2^2)=
\frac{r^2}{2} \delta \alpha(r) (\sigma_+^2+\sigma_-^2)\ .
\label{da_pert}
\end{equation}
This perturbation is $SU(2)$-invariant, 
but the first and second terms in the last expression have the $U(1)$-charges $+2$ and $-2$, respectively.
Other components of $SU(2)$-invariant perturbations cannot have $U(1)$-charges $\pm2$. 
(For example, $dt^2$ and $dt\sigma_+$ have the $U(1)$-charges $0$ and $+1$, respectively. 
Only $\sigma_\pm^2$ can have the charges $\pm2$.)
As a result, we obtain a decoupled perturbation equation for $\delta \alpha$,
\begin{equation}
\delta \alpha'' + \left( \frac{g'}{g}+\frac{3+5r^2}{r(1+r^2)} \right) \delta \alpha' 
+ \frac{8}{(1+r^2)g}\left(\frac{\beta-2}{r^2\beta} +\frac{2\beta h^2}{(1+r^2)g}\right) \delta \alpha = 0 \ .
\label{da_eq}
\end{equation}
The perturbation in the rotating frame \eqref{da_pert} is $\tau$-independent, but 
its expression is explicitly time periodic in the non-rotating frame: 
\begin{equation}
\delta g_{\mu\nu}dx^\mu dx^\nu=r^2 \delta \alpha(r) (e^{4i\Omega t}\bar{\sigma}_+^2+e^{-4i\Omega t}\bar{\sigma}_-^2)/2.
\label{da_pert_sigmabar}
\end{equation}
This is a normal mode whose frequency is given by $\omega= \pm 4\Omega$.
This corresponds to the onset of a ``$(J,M,K)=(0,0,\pm 2)$'' superradiant instability~\cite{Murata:2008xr}.

\begin{figure}[t]
\centering
\includegraphics[scale=0.45]{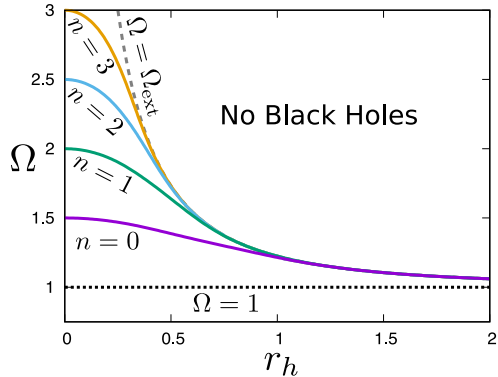}
\caption{Onset of the superradiant instability of MPAdS$_5$. 
The frequency of the extreme MPAdS$_5$ $\Omega=\Omega_\mathrm{ext}$ is shown in a dashed line; there are no regular MPAdS$_5$ in the upper right region. We also plot $\Omega=1$ in a dotted line which $\Omega_\mathrm{ext}$ asymptotes to in $r_h \to \infty$.
}
\label{fig:linear_reso_instability}
\end{figure}

We numerically search the critical values of $\Omega$ 
when the equation \eqref{da_eq} is solved by a nontrivial $\delta \alpha$ with trivial boundary conditions $\delta \alpha(r_h) = \delta \alpha(\infty) = 0$.
We can find not only the fundamental normal mode but also its overtones, and we label the modes by $n=0,1,2,\cdots$.
In Fig.~\ref{fig:linear_reso_instability}, we plot the angular velocities at the onset of the instabilities 
for $n \le 3$ modes.
The extreme MPAdS$_5$ saturates $\Omega_\mathrm{ext}=\sqrt{1+1/(2 r_h^2)}$, and in the upper right region there are no black holes described by Eq.\eqref{KerrAdS}.
All curves of the instability frequencies approach the extreme black hole frequency as $r_h$ increases.
The higher overtones with
$n>3$, which are not shown in the figure, also approach the extremal frequency quickly.
The results for the $n=0$ mode coincide with~\cite{Murata:2008xr}.

\section{Cohomogeneity-1 geons and black resonators}
\label{sec:cohomo1}

In this section, generalizing Eq.\eqref{KerrAdS}, we introduce a metric ansatz for cohomogeneity-1 geons and black resonators in the five-dimensional asymptotically AdS spacetime.
We discuss the symmetries that the ansatz possesses and argue that the ansatz is designed for the spacetime with a helical Killing vector field.
We also derive the expressions for thermodynamic quantities.

\subsection{Metric ansatz}

In section \ref{SRMPAdS}, 
we found that there is a superradiant instability for the $SU(2)$-invariant perturbation (\ref{da_pert}).
Here we consider the nonlinear extension of the normal mode from the onset of the $SU(2)$-invariant instability.
As a nonlinear generalization of Eq.(\ref{da_pert}), 
we take the metric ansatz as 
\begin{multline}
 ds^2=-(1+r^2)f(r)d\tau^2 + \frac{dr^2}{(1+r^2)g(r)}\\+\frac{r^2}{4} \left[
\alpha(r)\sigma_1^2 + \frac{1}{\alpha(r)}\sigma_2^2 + \beta(r)(\sigma_3+2 h(r)d\tau)^2
\right]\ .
\label{metricanz}
\end{multline}
This spacetime is still cohomogeneity-1: The metric nontrivially depends only on the radial coordinate $r$.
We introduce a new function $\alpha(r)$ which deforms the $S^2$ base space.
The product of the coefficients of $\sigma_1^2$ and $\sigma_2^2$  are fixed 
by the redefinition of the radial coordinate $r$.
The $U(1)$-symmetry that exists in the MPAdS$_5$ with equal angular momenta 
is broken in this ansatz unless $\alpha(r)$ is identically one.
It follows that the metric~(\ref{metricanz}) only has the $R_\tau\times SU(2)$ symmetries.
If we set $\alpha(r)=1$ trivially, 
our metric ansatz reduces to that for a MPAdS$_5$ black hole with equal angular momenta~(\ref{KerrAdS}).

How does the metric~(\ref{metricanz}) describe the black resonator and geon?
Let us consider asymptotically AdS spacetime: 
\begin{equation}
 f(r)\to 1,\quad 
 \alpha(r)\to 1,\quad 
 \beta(r)\to 1 \quad  (r\to \infty)\ .
\label{AsympAdS}
\end{equation}
Then $g(r)\to 1$ is automatically satisfied because of the Einstein equation.
Meanwhile, the above condition does not restrict the asymptotic value of $h(r)$.
We denote it by $\Omega \equiv h(\infty)$.
Under the asymptotically AdS condition~(\ref{AsympAdS}), 
the metric near the infinity 
becomes the same form as Eq.(\ref{MPAdSasym}).
The ansatz \eqref{metricanz} indeed corresponds to the rotating frame: 
Later, we will check that $\partial_\tau$ is normal to the horizon (i.e. $h(r)$ is zero at the horizon), 
but here we assume this.

Changing coordinates by Eq.(\ref{psishift}), we can go to the non-rotating frame.
However, we cannot simply shift the value of $h(\infty)$ while keeping the form of the metric~(\ref{metricanz}).
In the non-rotating frame, actually, there appears explicit periodic time dependence in the metric.
This can be seen by rewriting the corresponding part of the metric with $(t,\psi)$ as
\begin{equation}
\alpha \sigma_1^2 + \frac{1}{\alpha}\sigma_2^2 
= 
2\left(\alpha+\frac{1}{\alpha}\right)\bar{\sigma}_+ \bar{\sigma}_- 
+\left(\alpha-\frac{1}{\alpha}\right)(e^{4i\Omega t}\bar{\sigma}_+^2+e^{-4i\Omega t}\bar{\sigma}_-^2)\ .
\label{tdeppart}
\end{equation}
The second part in the right hand side is a nonlinear generalization of Eq.\eqref{da_pert_sigmabar}.

As a consequence, the asymptotic time translation $\partial_t$ and rotation $\partial_\psi$ 
are {\textit{not}} independently Killing vectors of the whole spacetime~(\ref{metricanz}) unlike the MPAdS$_5$.
Only an appropriate combination of $\partial_t$ and $\partial_\psi$ gives a Killing vector as follows:
\begin{equation}
K\equiv \frac{\partial}{\partial \tau} =  \frac{\partial}{\partial t}+\Omega\, \frac{\partial}{\partial (\psi/2)}\ .
\label{helical_killing_vector}
\end{equation}
This Killing vector is ``helical'' with respect to the asymptotically non-rotating frame $(t,\psi)$. 
The asymptotic behavior of the norm of the Killing vector is
\begin{equation}
 g_{\mu\nu}K^\mu K^\nu=g_{\tau\tau} \to -r^2(1-\Omega)\qquad (r\to \infty)\ .
\end{equation}
For $\Omega>1$, the Killing vector is asymptotically spacelike.
Therefore, the solutions with $\Omega>1$ and $\alpha(r)\neq 1$ express dynamical spacetime.
In the following sections, we will construct such solutions. 
From Eq.\eqref{tdeppart}, we find that the dynamical spacetime is time periodic with respect to the asymptotic time $t$.
The solutions without a horizon are geons~\cite{Horowitz:2014hja}.
The solutions with horizons are called black resonators~\cite{Dias:2015rxy}.
The cohomogeneity-1 metric~(\ref{metricanz}) is a simple ansatz for studying geons and black resonators.

Because of the cohomogeneity-1 ansatz, 
the Einstein equations reduce to a set of ordinary differential equations:
\begin{align}
\begin{split}
f'=&\frac{1}{r(1+r^2)^2 g \alpha^2 (r\beta'+6\beta)}
[
4 r^2 h^2 (\alpha^2-1)^2 \beta\\
&+r (r^2+1) g\{r(1+r^2)  f \alpha'{}^2 \beta  
-r^3   h'{}^2 \alpha^2 \beta^2 
-2 (2+3 r^2)  f \alpha^2   \beta'\}\\
&
-4 (1+r^2) f  \{6 r^2 \alpha^2 \beta (g-1)+3 g \alpha^2 \beta + (\alpha^2-\alpha \beta+1)^2-4 \alpha^2\}
]
\label{EOMf}
\end{split}
\ ,\\
\begin{split}
g'=&\frac{1}{6 r(1+r^2)^2 f \alpha^2 \beta}
[
-4 r^2 h^2   (\alpha^2-1)^2\beta\\
&+r(1+r^2)g\{
-r (1+r^2)  f \alpha'{}^2 \beta
+r^3 h'{}^2 \alpha^2 \beta^2\\
&- (-r(1+r^2) f'+2 f) \alpha^2 \beta'\}
+4 (1+r^2) f  \{-6r^2\alpha^2 \beta(g - 1)-3 g \alpha^2 \beta \\
& \hspace{28ex} 
+\alpha^4+4 \alpha^3 \beta-5 \alpha^2 \beta^2-2 \alpha^2+4 \alpha \beta+1\}
]
\label{EOMg}
\end{split}
\ ,\\
\begin{split}
h''=&\frac{1}{2 r^2 (1+r^2) \alpha^2 \beta f g}
[
8 f h (\alpha^2-1)^2\\
&-r (1+r^2) h' \alpha^2 \{r (f g' \beta -f' g \beta +3 f g \beta')+10 f g \beta \} 
]
\label{EOMh}
\end{split}
\ ,\\
\begin{split}
\alpha''=&\frac{1}{2 r^2 (1+r^2)^2 f \alpha g \beta}
[
2 r^2 (r^2+1)^2  f g \alpha'{}^2 \beta\\
&-r (r^2+1) \alpha \alpha'  \{r(1+r^2)(f g \beta)' +2 (3+5 r^2)  f g \beta\}\\
&-8 (\alpha^2-1) \{
r^2 h^2 \beta (\alpha^2 + 1)-(1+r^2) f \alpha(\alpha-\beta) -(1+r^2)f
\}
]
\label{EOMa}
\end{split}
\ ,\\
\begin{split}
\beta''=&\frac{1}{(2 r^2 (1+r^2)) f g \alpha^2 \beta}
[
-2r^4  g h'{}^2 \alpha^2 \beta^3 \\
&-r \alpha^2 \beta' \{
r(1+r^2)(f' g \beta+ f g' \beta- f g \beta')
+2 (3+5 r^2)f g \beta \}\\
&-8 f \beta (\alpha^4+\alpha^3 \beta-2 \alpha^2 \beta^2-2 \alpha^2+\alpha \beta+1)
]
\label{EOMb}
\end{split}
\ .
\end{align}
We will solve them numerically.
In the case of four-dimensional AdS geons and black resonators, 
three-dimensional partial differential equations were solved~\cite{Horowitz:2014hja,Dias:2015rxy}.
In contrast, considering five-dimensions and taking a special ansatz as in Eq.(\ref{metricanz}), 
we can reduce the problem to solving one-dimensional differential equations.

\subsection{Boundary stress tensor and thermodynamic variables}

The thermodynamic variables for our spacetime can be constructed from the data on the AdS boundary and horizon.
Solving the equations of motion~(\ref{EOMf}-\ref{EOMb}) near the AdS boundary, 
we obtain the asymptotic solution of the metric components as
\begin{equation}
\begin{split}
&f(r)=1+\frac{c_f}{r^4}+\cdots\ ,\quad
g(r)=1+\frac{c_f+c_\beta}{r^4}+\cdots\ ,\\
&h(r)=\Omega+\frac{c_h}{r^4}+\cdots\ ,\quad
\alpha(r)=1+\frac{c_\alpha}{r^4}+\cdots\ ,\quad
\beta(r)=1+\frac{c_\beta}{r^4}+\cdots\ ,
\end{split}
\label{asym}
\end{equation}
where $c_f, c_h, c_\alpha$, and $c_\beta$ are the constants undetermined in the series solution.

We can employ the Ashtekar-Das method to construct conserved charges \cite{Ashtekar:1999jx}.
The stress energy tensor on the AdS boundary can be given by \cite{Kinoshita:2008dq}\footnote{%
This is equivalent to the holographic stress energy tensor derived by carrying out holographic renormalization~\cite{Balasubramanian:1999re,deHaro:2000vlm} 
up to the Casimir contribution which is absent in Eq.\eqref{TijFromWeyl}. 
}
\begin{equation}
 8\pi G_5 T_{ij}=-\frac{r^2}{2}C_{i\rho j \sigma} n^\rho n^\sigma\bigg|_{r=\infty}\ ,
 \label{TijFromWeyl}
\end{equation}
where $i,j$ run over the coordinates on the AdS boundary, 
$n^\mu$ is the unit normal to a bulk $r$-constant surface, 
and $C_{\mu\nu\rho\sigma}$ is the bulk Weyl tensor.
Substituting the boundary expansion (\ref{asym}) into the above expression, we obtain
\begin{multline}
 8\pi G_5 T_{ij}dx^i dx^j
= \frac{1}{2}(c_\beta-3c_f) d\tau^2
+2 c_h d\tau(\sigma_3+2\Omega d\tau)
-\frac{c_f+c_\beta}{8}(\sigma_1^2+\sigma_2^2)\\
+\frac{c_\alpha}{2}(\sigma_1^2-\sigma_2^2)
+\frac{1}{8}(-c_f+3c_\beta)(\sigma_3+2\Omega d\tau)^2
\ .
\end{multline}
In the non-rotating frame $(t,\psi)$, the boundary stress tensor is rewritten as
\begin{multline}
 8\pi G_5 T_{ij}dx^i dx^j
= \frac{1}{2}(c_\beta-3c_f) dt^2
+2 c_h dt \bar{\sigma}_3-\frac{c_f+c_\beta}{8}(\bar{\sigma}_1^2+\bar{\sigma}_2^2)\\
+c_\alpha (e^{4i\Omega t}\bar{\sigma}_+^2+e^{-4i\Omega t}\bar{\sigma}_-^2)
+\frac{1}{8}(-c_f+3c_\beta)\bar{\sigma}_3^2
\ .
\label{Tmunu_non-rot}
\end{multline}
The energy density ($\propto T_{tt}$) and angular momentum density ($\propto T_{t\psi}$) 
depend on neither time nor spatial coordinates 
unlike the case of the four-dimensional geons and black resonators constructed in~\cite{Horowitz:2014hja,Dias:2015rxy}.
Only the stress part (the coefficients of $\bar{\sigma}_{\pm}$) depends on the asymptotic time $t$.
The energy and angular momentum are given by\footnote{%
We also define the angular momentum with respect to $\psi/2\in [0,2\pi)$.
}
\begin{equation}
E=\int d\Omega_3 T_{tt} =\frac{\pi(c_\beta-3c_f)}{8 G_5}\ ,\quad
J=-\int d\Omega_3 T_{t (\psi/2)} =-\frac{\pi c_h}{2 G_5}\ .
\label{EJdef}
\end{equation}

The entropy $S$ and temperature $T$ can be defined from the horizon data as
\begin{equation}
S=\frac{\pi^2 r_h^3 \sqrt{\beta(r_h)}}{2G_5}\ ,\quad
T=\frac{(1+r_h^2)\sqrt{f'(r_h)g'(r_h)}}{4\pi}\ .
\end{equation}
For these thermodynamic quantities, the first law of thermodynamics is
\begin{equation}
dE = T dS + \Omega dJ \ .
\end{equation}
In the case of the MPAdS$_5$, where we obtain Eq.\eqref{OmegaMPAdS} and Eq.\eqref{mpads_thermo}, 
we can analytically check that this relation is satisfied~\cite{Gibbons:2004ai}.\footnote{%
The normalization of our $J$ is the same as that in \cite{Kunduri:2006qa}, and it is related to the equal angular momentum case of $J_{a,b}$ in \cite{Gibbons:2004ai} as $J=J_a+J_b$ with $a=b$.
}
In the numerical solutions of the geons and black resonators we will construct in the following sections, 
we check that the first law of thermodynamics is satisfied within numerical accuracy.
For notational simplicity, we will set $G_5=1$ hereafter.
We can easily recover the dependence on $G_5$ by
$E\to G_5 E$, $J\to G_5 J$, and $S\to G_5 S$.

\section{Geons}
\label{sec:geon}

Geons appear nonlinearly from a normal mode of global AdS \cite{Dias:2011ss,Horowitz:2014hja}.
For them, which are horizonless, the boundary condition at the center $r=0$ of the global AdS 
is different from that for the black resonators.
In this section, therefore, we treat the geons separately.
We firstly consider the perturbative construction of the geons.
We then compute full nonlinear solutions numerically.

\subsection{Perturbative construction}

Let us start from a linear perturbation of global AdS.
The perturbation equation takes the same form as Eq.(\ref{da_eq}) where,
as the background, we take the global AdS in a rotating frame at infinity: $f(r)=g(r)=\alpha(r)=\beta(r)=1$ and $h(r)=\Omega$. 
Then, the perturbation equation can be solved analytically by
\begin{equation}
\delta \alpha = r^2(1+r^2)^{-2\Omega}{}_2F_1(1-2\Omega,3-2\Omega;4;-r^2)\ .
\label{da_exact}
\end{equation}
At the AdS boundary, this behaves
\begin{equation}
\delta \alpha \to \frac{6}{\Gamma(3-2\Omega)\Gamma(3+2\Omega)}\quad (r\to \infty)\ .
\end{equation}
Therefore, $\delta \alpha$ satisfies $\delta \alpha\to 0$ in $r\to\infty$ only if $\Omega=(3+n)/2$ ($n=0,1,2,\cdots$).
These corresponds to a tower of gravitational normal modes in the global AdS.
The values at $r_h \to 0$ in Fig.~\ref{fig:linear_reso_instability} agree with these analytical results.
Geons appear from these normal modes as nonlinear gravitational solutions.

We continue the perturbative expansion to higher orders and construct geons perturbatively in a way 
similar to~\cite{Dias:2011ss,Horowitz:2014hja,Martinon:2017uyo,Fodor:2017spc}. (See appendix~\ref{pGeon} for details.)
Up to the fourth order in a small parameter $\epsilon$, 
we obtain the perturbative expressions of $E$, $J$, and $\Omega$ for the geons as
\begin{equation}
\begin{split}
E&=\frac{\pi}{8}\left(\frac{3}{10}\epsilon^2+\frac{803}{84000}\epsilon^4\right)\ ,\quad
J=\frac{\pi}{8}\left(\frac{1}{5}\epsilon^2+\frac{2549}{378000}\epsilon^4\right)\ ,\\
\Omega&=\frac{3}{2}-\frac{1}{180}\epsilon^2-\frac{356}{2338875}\epsilon^4\ .
\end{split}
\label{EJOmega_pert}
\end{equation}
One can check that the first law of thermodynamics is satisfied: 
$ dE/dJ = (dE/d\epsilon)/(dJ/d\epsilon)=\Omega + \mathcal{O}(\epsilon^4)$.

\subsection{Numerical construction of the full geon solutions}

\begin{figure}
  \centering
  \subfigure
 {\includegraphics[scale=0.33]{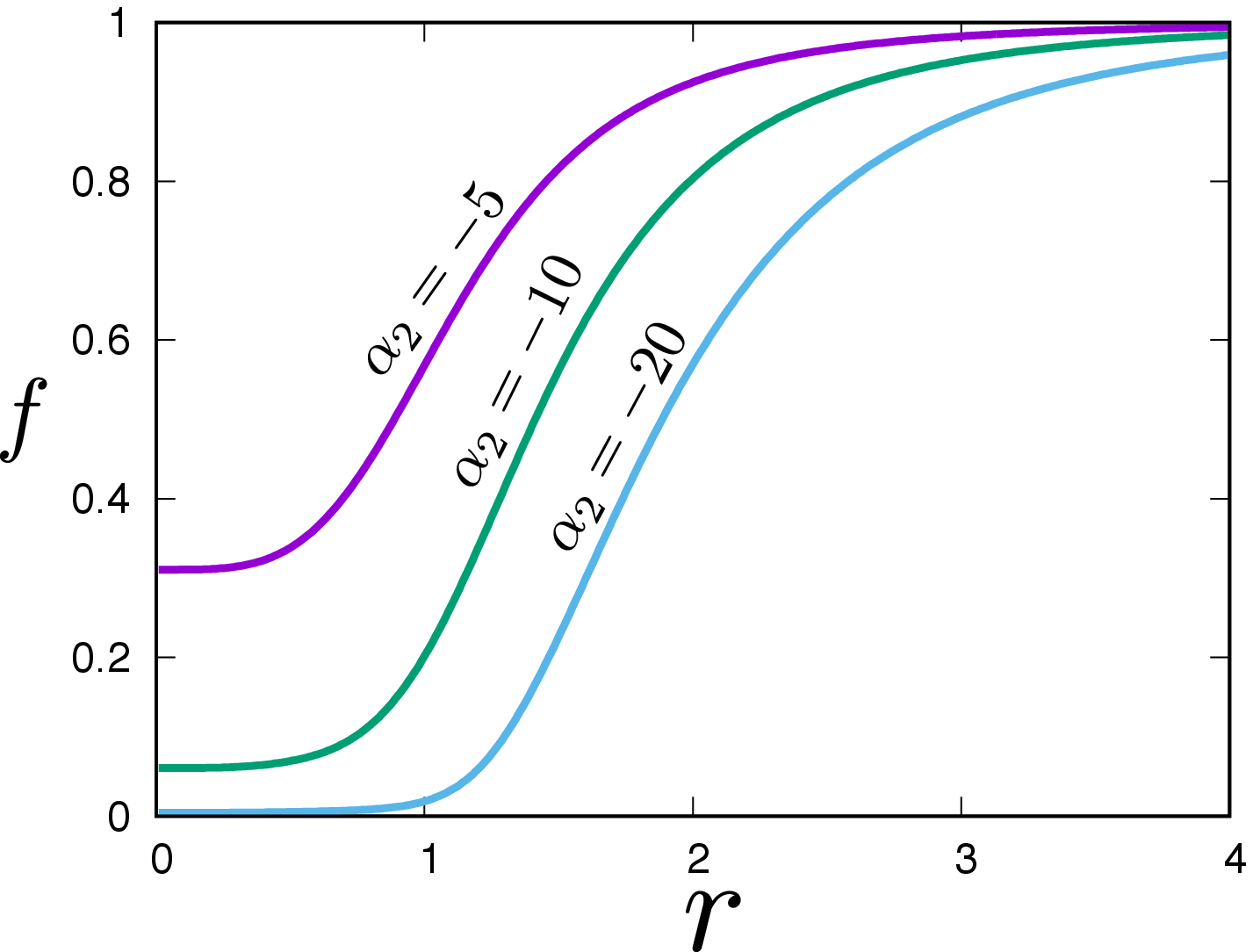}
  }
\subfigure
 {\includegraphics[scale=0.33]{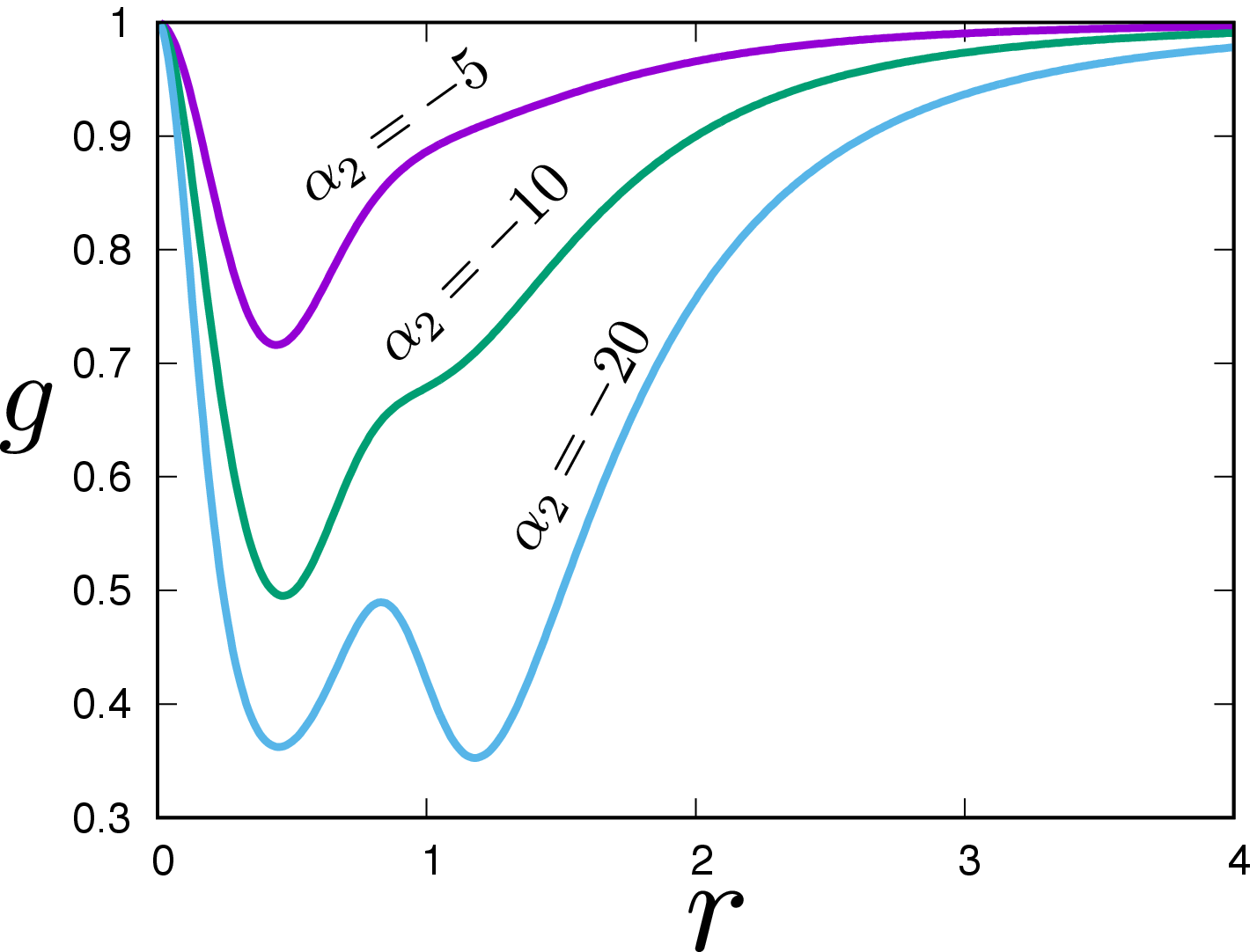}
  }
\subfigure
 {\includegraphics[scale=0.33]{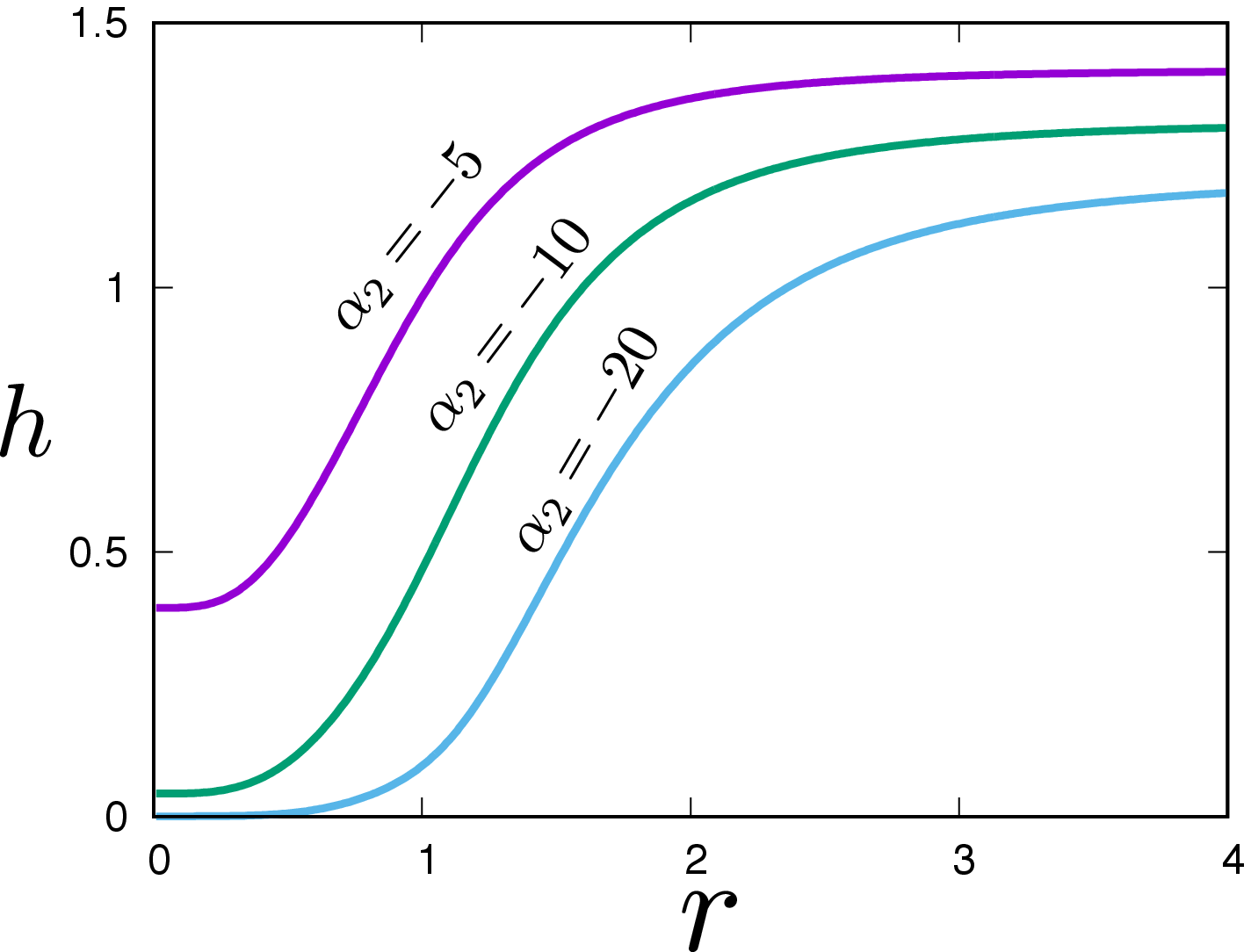}
  }
\subfigure
 {\includegraphics[scale=0.33]{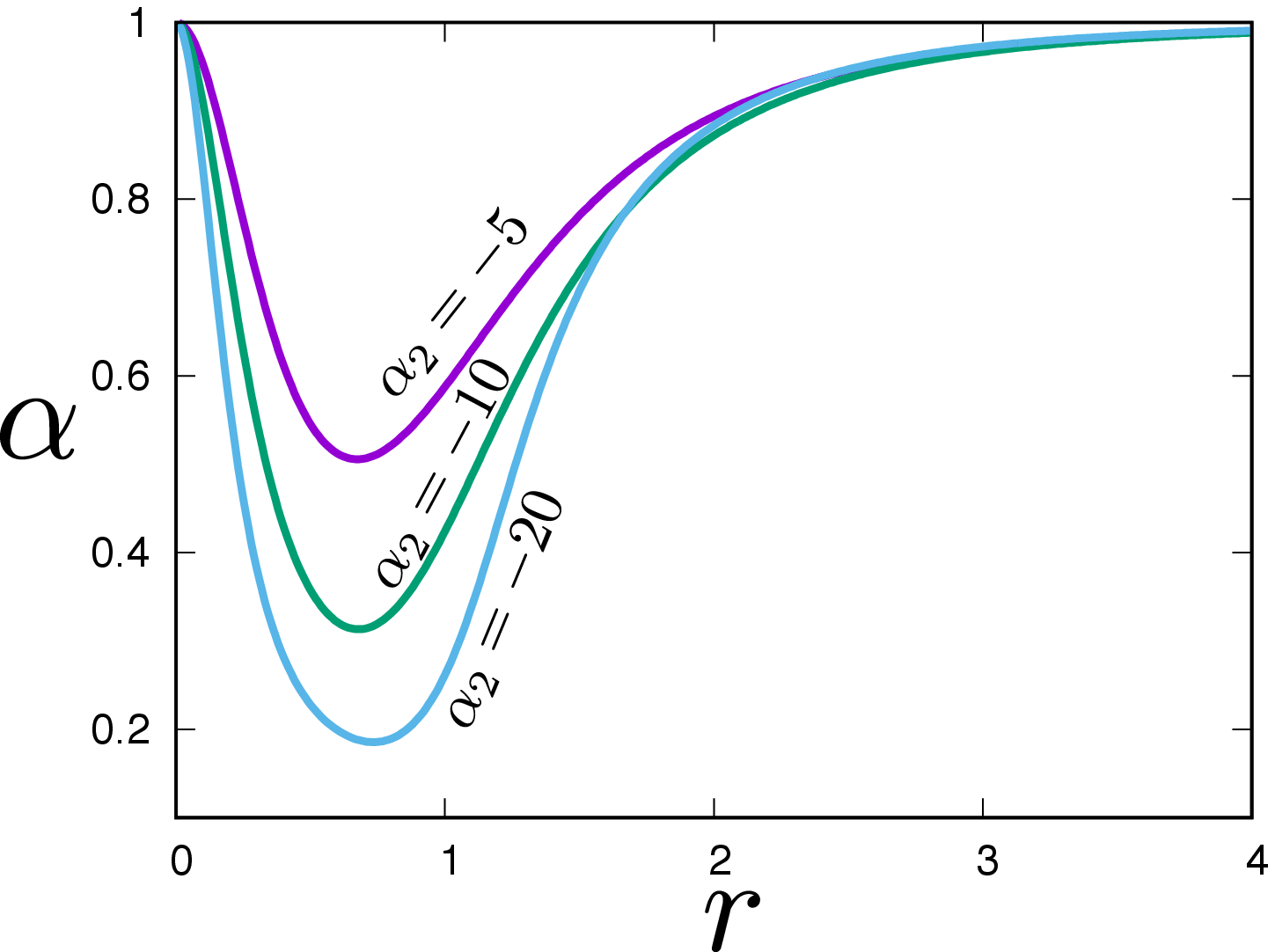}
  }
\subfigure
 {\includegraphics[scale=0.33]{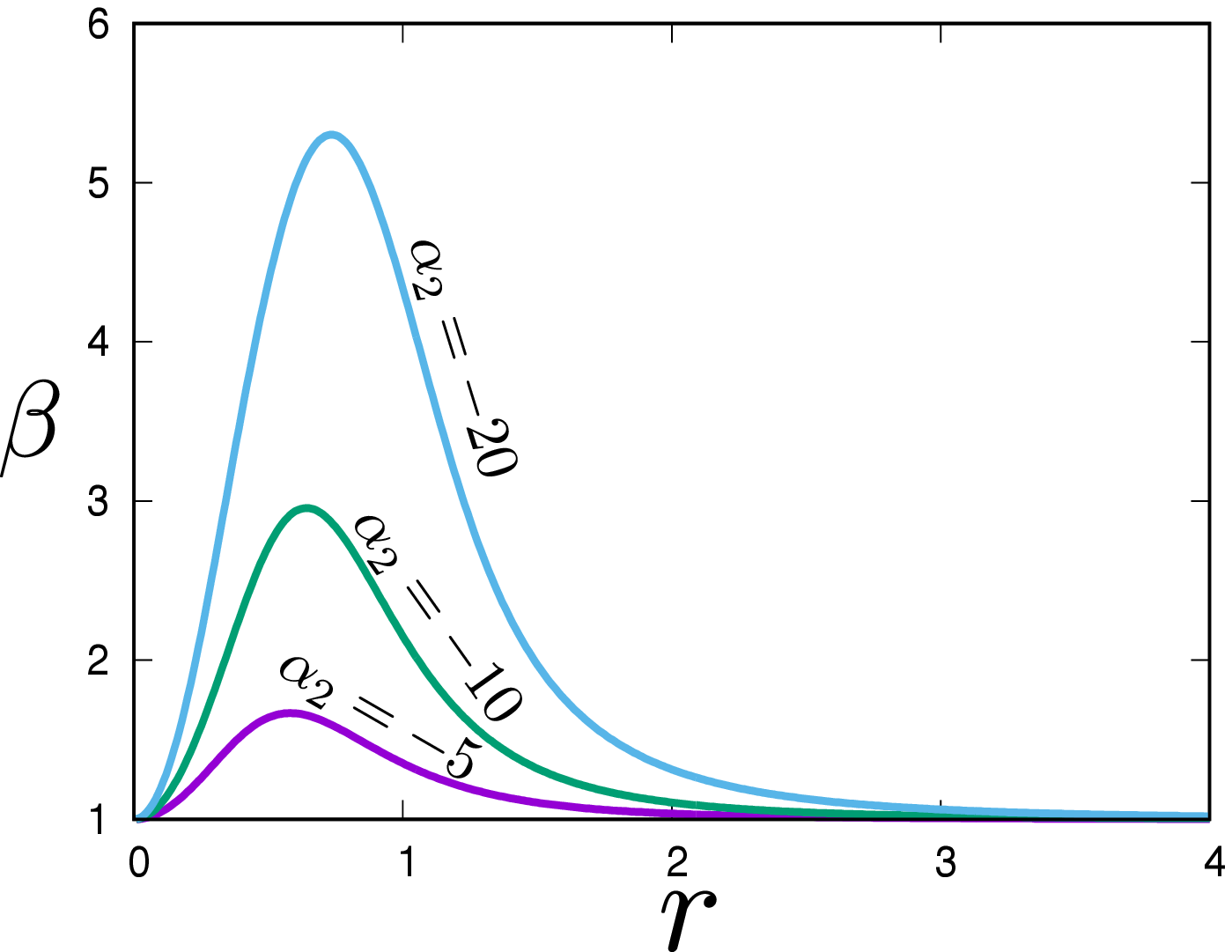}
  }
\subfigure
 {\includegraphics[scale=0.33]{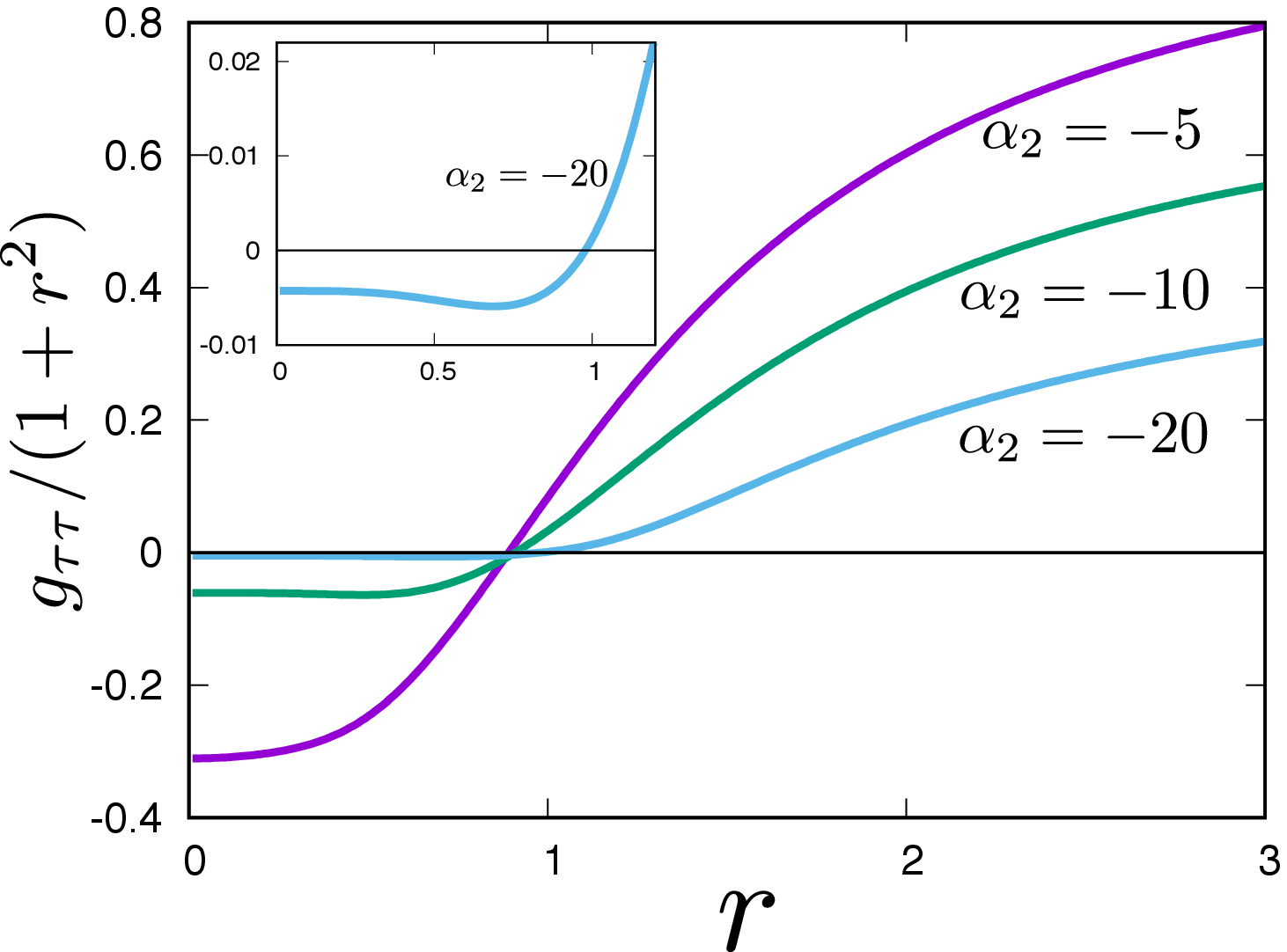}
  }
\caption{%
The metric components of the geons when $\alpha_2=-5,-10,-20$.
In the bottom-right, the norm of the Killing vector \eqref{helical_killing_vector} is plotted.
\label{geon_sol} 
}
\end{figure}

To avoid a conical singularity at the center of the AdS $r=0$, 
we impose a regular boundary condition there: $g(0)=1$, $\alpha(0)=1$ and $\beta(0)=1$. 
Solving the equations of motion around $r=0$, we obtain series solutions
\begin{equation}
\begin{split}
&f(r)=f_0 + \mathcal{O}(r^4)\ ,\quad 
g(r)=1-\beta_2 r^2 + \mathcal{O}(r^4)\ ,\quad
h(r)=h_0 + \mathcal{O}(r^4)\ ,\\
&\alpha(r)=1 + \alpha_2 r^2 + \mathcal{O}(r^4)\ ,\quad
\beta(r)=1+\beta_2 r^2 + \mathcal{O}(r^4)\ .
\end{split}
\end{equation}
We have four free parameters, $f_0,\,h_0,\,\alpha_2,$ and $\beta_2$.
When $\alpha_2=0$, the solution is simply the global AdS.
To construct a geon solution, we turn on $\alpha_2\neq 0$ as the parameter and tune the other three parameters $f_0$, $h_0$, and $\beta_2$ so that 
the asymptotically AdS conditions~(\ref{AsympAdS}) are satisfied.\footnote{%
We start from a normal mode for the global AdS, $f_0=1$, $h_0=3/2$, and $\beta_2=0$, and turn on a small $\alpha_2$.
Once we succeed in finding a solution, we vary $\alpha_2$ slightly while using 
the previous result of ($f_0,h_0,\beta_2$) as an initial guess for a next solution.
}
Without loss of generality, 
we can assume $\alpha_2<0$ because 
the field redefinition
\begin{equation}
 \alpha(r)\to \frac{1}{\alpha(r)}
\label{isomet}
\end{equation}
just exchanges the roles of $\sigma_1$ and $\sigma_2$.

In Fig.~\ref{geon_sol}, the numerical geon solutions are shown for $\alpha_2=-5,-10,-20$.\footnote{%
We consider the fundamental ($n=0$) tone unless specified.
}
Their corresponding thermodynamical quantities are $(E,J,\Omega)=$(2.11, 1.45, 1.41), (5.84, 4.20, 1.31), (15.3, 11.8, 1.21), respectively.
The metric is clearly deformed from the pure AdS as $|\alpha_2|$ grows.
In particular, the function $\alpha(r)$ measures the deformation of the base space $S^2$.
We find a dip in the profile of $\alpha(r)$ around $r\simeq 0.8$. This indicates that 
the deformation of $S^2$ is localized around there.
In the bottom-right panel of Fig.~\ref{geon_sol}, we also show the norm of the Killing vector: 
$g_{\mu\nu}K^\mu K^\nu=g_{\tau\tau}$. For visibility, we normalize it by $1+r^2$.
Near the center of the AdS, $K$ is timelike, i.e.,~the spacetime is stationary in this region.
Far from the center ($r \gtrsim 1$), however, it becomes spacelike, and the spacetime is dynamical.

\begin{figure}
  \centering
  \subfigure[$E$ vs $J$]
 {\includegraphics[scale=0.45]{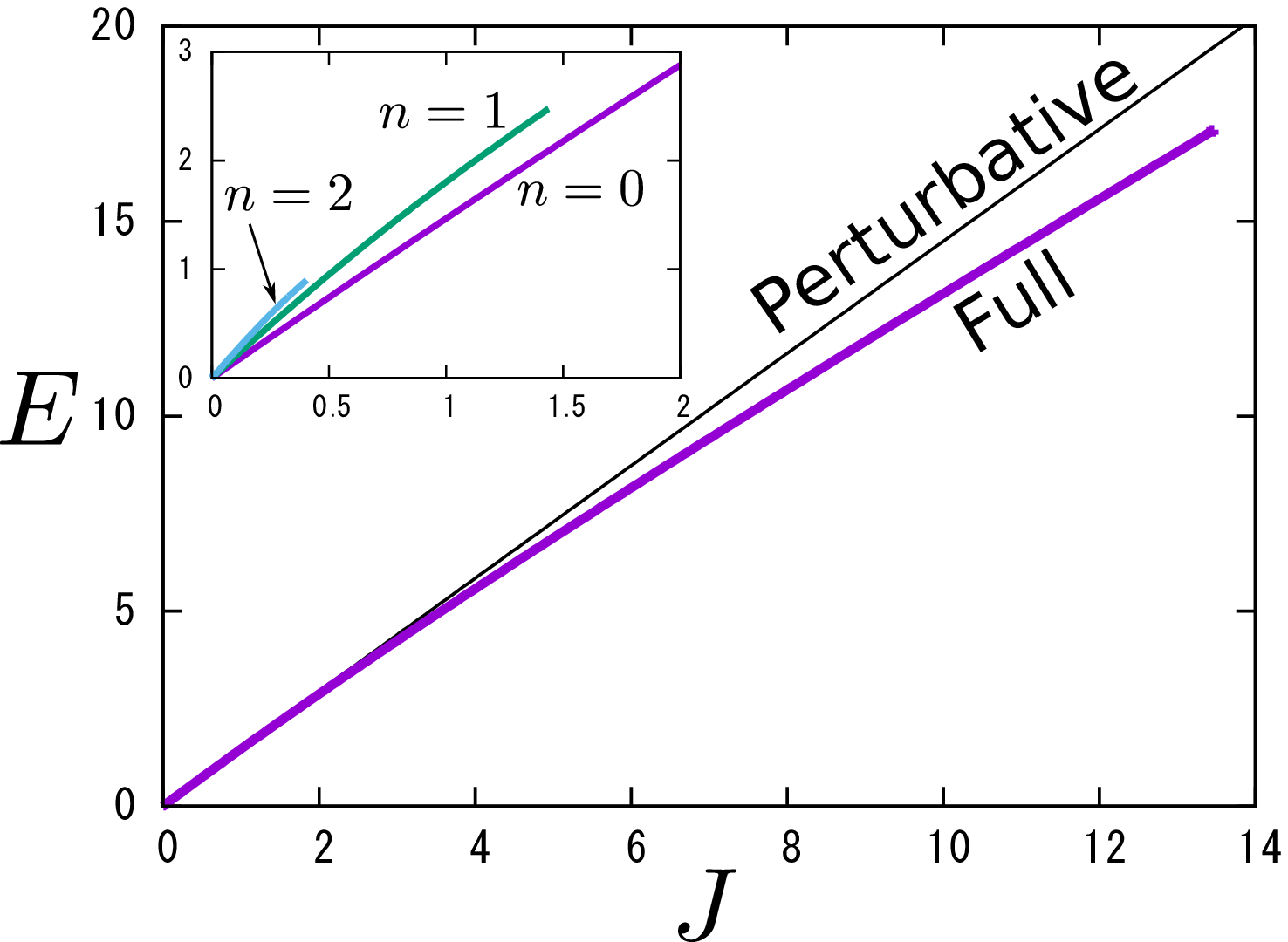}\label{EvsJ}
  }
  \subfigure[$\Omega$ vs $J$]
 {\includegraphics[scale=0.45]{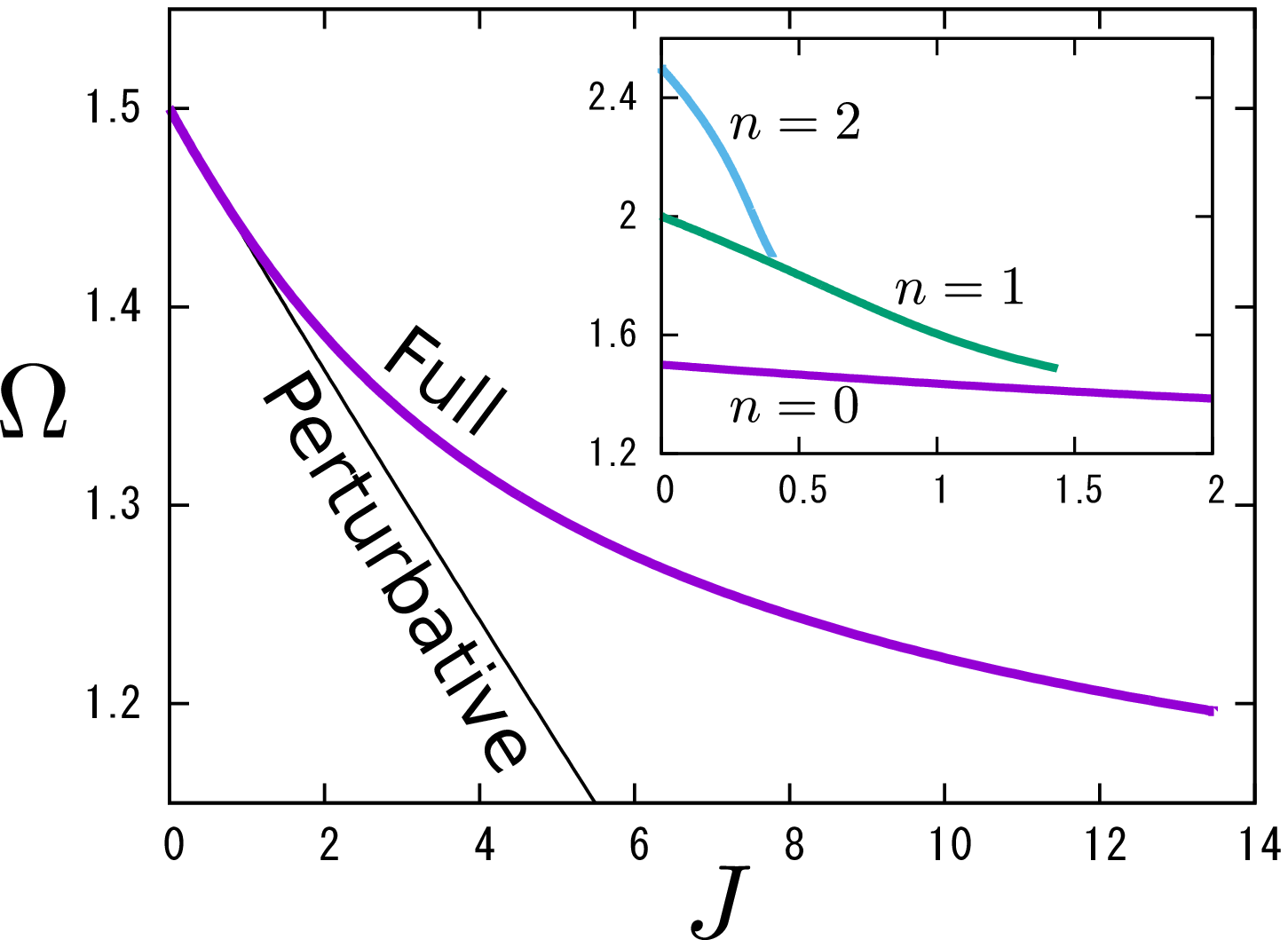}\label{OmvsJ}
  }
 \caption{%
 Mass $E$, angular momentum $J$ and angular velocity $\Omega$ for the family of geons.
$E$ and $\Omega$ are shown as functions of $J$.
Purple and black curves correspond to the numerically constructed full solutions and 
perturbative solutions, respectively.
Those for overtones $n=0,1,2$ are also shown in insets.
\label{geon_thermo} 
}
\end{figure}

In Fig.~\ref{geon_thermo}, we plot the mass $E$ and the angular velocity $\Omega$ as functions of the angular momentum $J$ 
and compare them between the full numerical and perturbative solutions,
which are shown in the purple and black curves, respectively.
For a small angular momentum $J \lesssim 1$, 
the perturbative results agree well with the numerical ones.
In $J\gtrsim 1$, however, the former deviates clearly from the latter.
We are able to obtain the geons even in a highly non-perturbative regime
thanks to the cohomogeneity-1 metric ansatz~(\ref{metricanz}).
We also checked the first law of thermodynamics, $dE/dJ=\Omega$, is satisfied within numerical accuracy.
In the insets of the figure, 
we also plot 
$E$ and $\Omega$ of the geons with the overtones $n=0,1,2$ 
as functions of angular momentum $J$. 
The geons with higher overtones have larger energy. 
Although the curves for $\Omega$ are likely to intersect each other, 
we could not find the intersection within our numerical limitations.

\section{Black resonators}
\label{sec:reso}

In this section, we construct the black resonators numerically by solving the ordinary differential equations~(\ref{EOMf}-\ref{EOMb}).
We calculate the thermodynamic quantities for the black resonators and construct the phase diagram for the black holes.
We also compare the thermodynamic quantities between the black resonators and MPAdS$_5$.

\subsection{Numerical solutions}

For black resonators, we impose that the metric functions $f(r)$ and $g(r)$ 
become zero at the horizon $r=r_h$. 
The asymptotic expansion near the horizon can be written in general as
\begin{equation}
\begin{split}
 &f(r)=f_1(r-r_h)+ \cdots\ ,\quad
  g(r)=g_1(r-r_h)+ \cdots\ ,\\
&h(r)=h_0+h_1(r-r_h)+\cdots\ ,\quad
\alpha(r)=\alpha_0+\alpha_1(r-r_h)+\cdots\ ,\\
&\beta(r)=\beta_0+\beta_1(r-r_h)+\cdots\ .
\end{split}
\end{equation}
Solving the equations of motion near the horizon, we find that $(f_1,h_1,\alpha_0,\beta_0)$ are not determined in the series expansion 
and the other coefficients are given in terms of them as
\begin{equation}
\begin{split}
&
 g_1=\frac{2(2r_h^2\alpha_0+\alpha_0^2-\alpha_0 \beta_0+1)}{\alpha_0 r_h (1+r_h^2)}\ ,\quad
\alpha_1=\frac{2(\alpha_0^2-1)(\alpha_0^2-\alpha_0 \beta_0+1)}{r_h \beta_0 (2r_h^2 \alpha_0+\alpha_0^2-\alpha_0 \beta_0+1)}\ ,\\
&h_0=0\ ,\quad\beta_1=-\frac{r_h^2 \beta_0^2 h_1^2}{f_1(1+r_h^2)}
-\frac{2(\alpha_0^4+\alpha_0^3 \beta_0-2 \alpha_0^2 \beta_0^2-2 \alpha_0^2+\alpha_0 \beta_0+1)}{
r_h \alpha_0 (2 r_h^2 \alpha_0+\alpha_0^2-\alpha_0 \beta_0+1)}\ .
\end{split}
\end{equation}
As the lowest order equation in $r-r_h$, we obtain $(\alpha_0-1)h_0=0$. 
Both $\alpha_0=1$ and $h_0=0$ can solve this equation.
The case of $\alpha_0=1$ corresponds to the MPAdS$_5$, where both $h_0 = 0$ and $h_0 \neq 0$ are allowed as we considered in section~\ref{sec:frames}.
If $\alpha_0 \neq 1$, we can construct the black resonator, but in this case we need $h_0 = 0$.
We choose the latter here.
Therefore, $K$ is the null generator of the horizon in Eq.\eqref{metricanz}.
There are five free parameters at the horizon: $(r_h,f_1,h_1,\alpha_0,\beta_0)$.
Three of them can be fixed by matching them with the asymptotically AdS condition~(\ref{AsympAdS}).
Therefore, the black resonators are in a two parameter family.
In our numerical calculations, we specify $(r_h, \alpha_0)$ as the parameters and tune $(f_1,h_1,\beta_0)$ so that Eq.(\ref{AsympAdS}) is satisfied.
For a given $r_h$, we start from a MPAdS solution at the onset of a superradiant instability and turn on a small deformation of $\alpha_0$ from 1.
Once a black resonator solution is successfully obtained, we use it as an initial guess for a next solution where $\alpha_0$ is slightly varied.
Without loss of generality, we assume $\alpha_0<1$ because of Eq.(\ref{isomet}).

\begin{figure}
  \centering
  \subfigure
 {\includegraphics[scale=0.33]{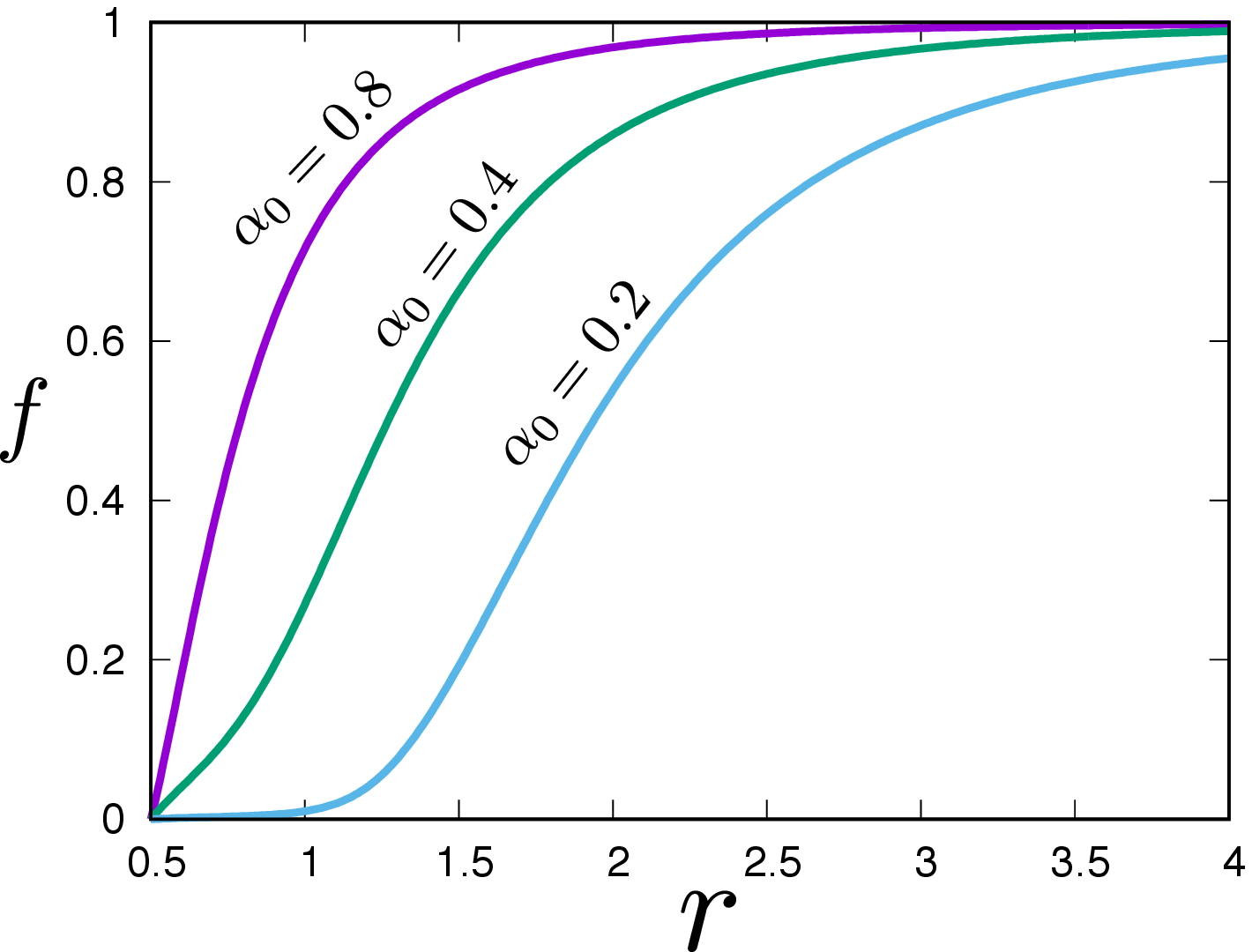}
  }
\subfigure
 {\includegraphics[scale=0.33]{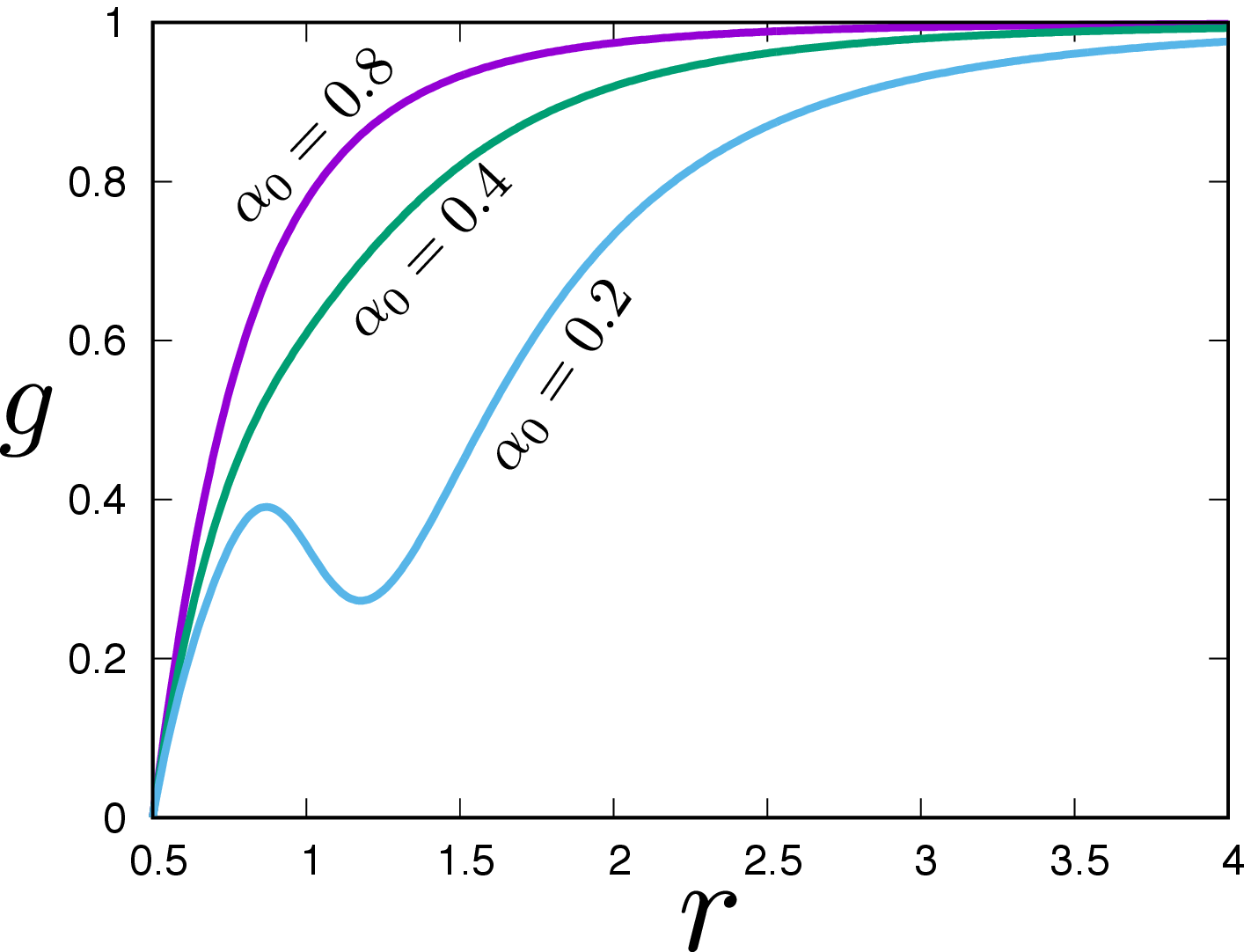}
  }
\subfigure
 {\includegraphics[scale=0.33]{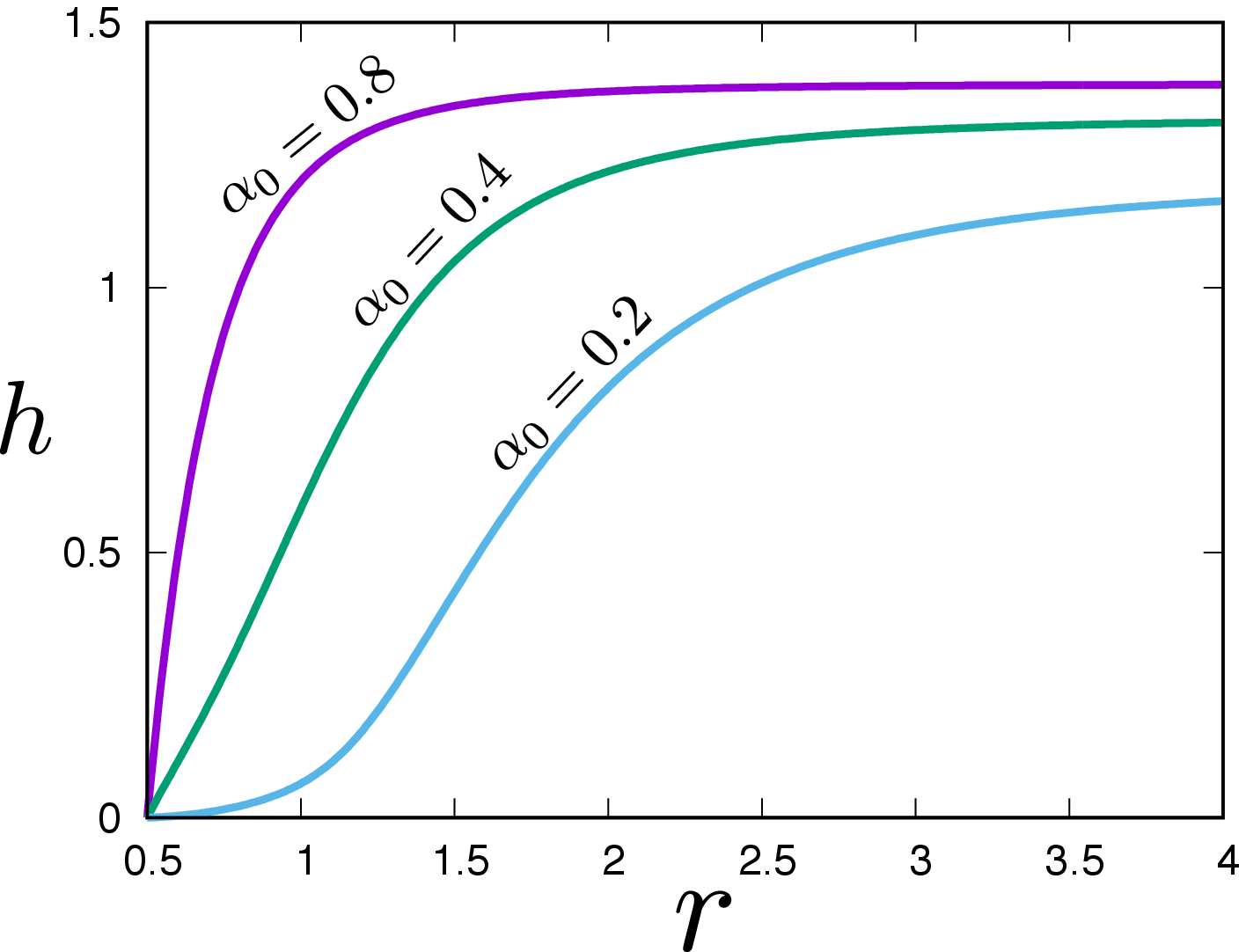}
  }
\subfigure
 {\includegraphics[scale=0.33]{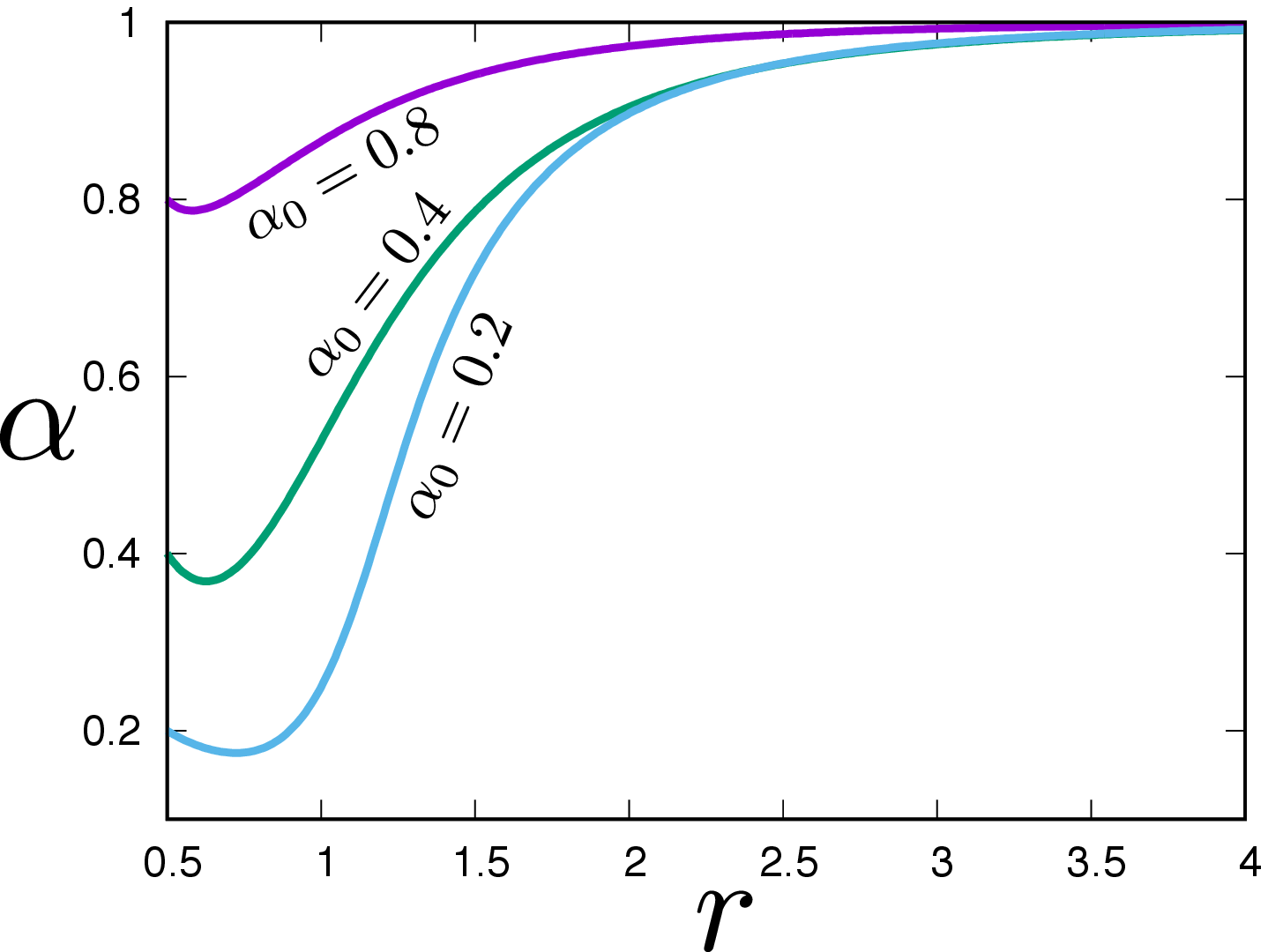}
  }
\subfigure
 {\includegraphics[scale=0.33]{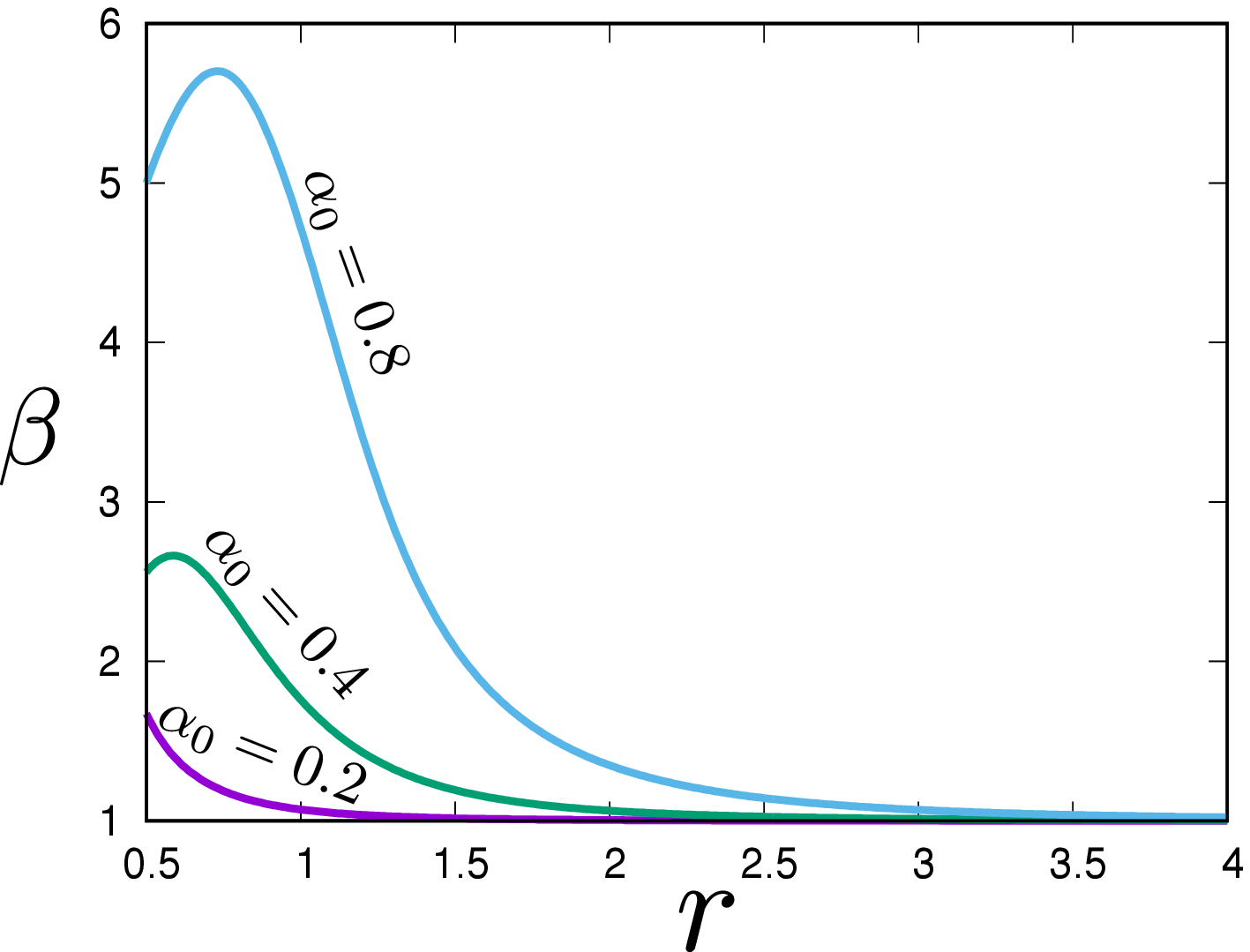}
  }
\subfigure
 {\includegraphics[scale=0.33]{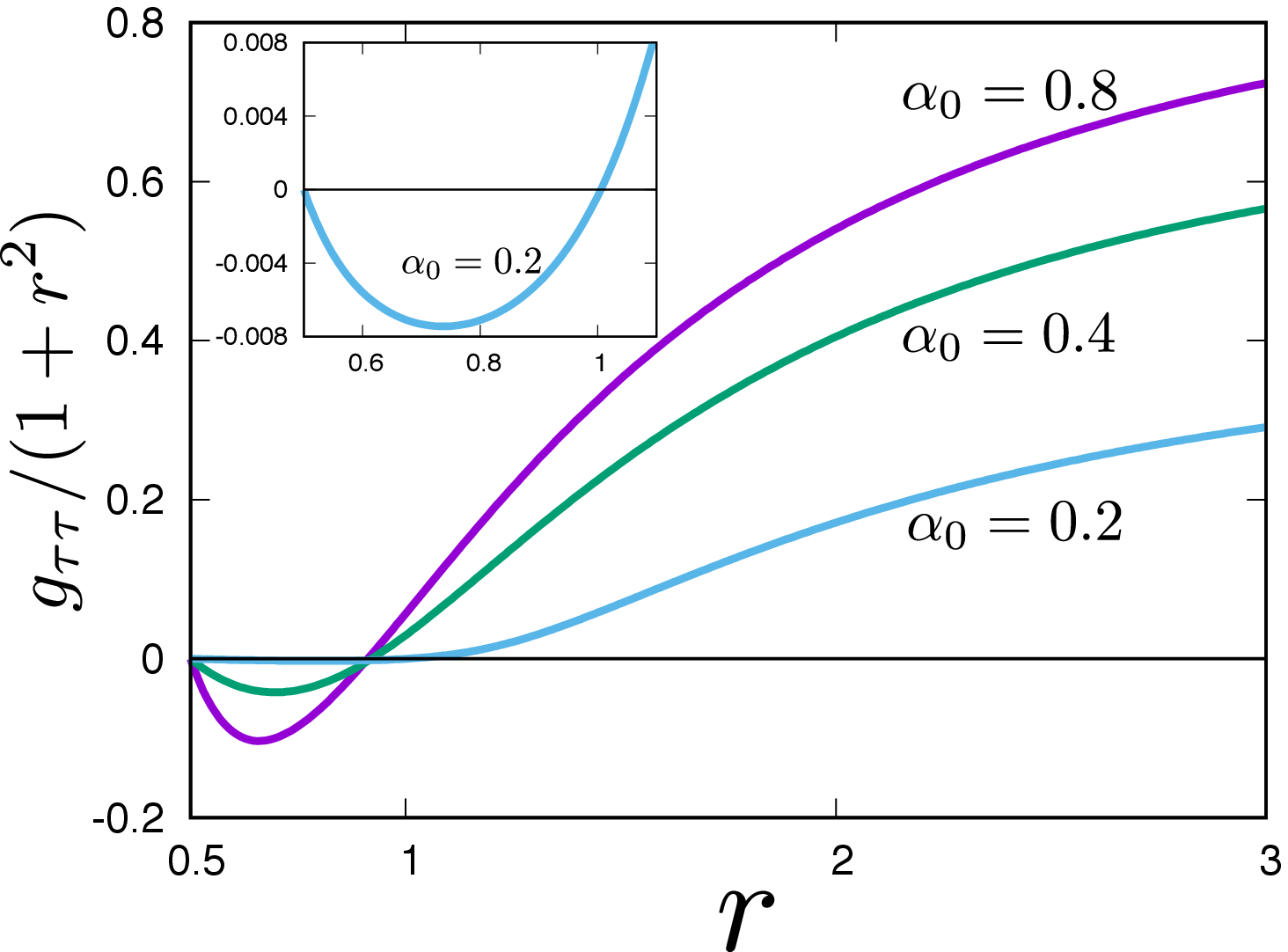}
  }
\caption{The metric components for the black resonators with $r_h=0.5$ and $\alpha_0=0.8,\,0.4,\,0.2$.
In the bottom-right panel, the norm of the Killing vector $K$ is plotted. 
\label{BR_sol}
}
\end{figure}

In Fig.~\ref{BR_sol}, we show the metric components for the black resonator solutions with $r_h=0.5$ and $\alpha_0=0.8,\,0.4,\,0.2$.
Their corresponding thermodynamical quantities are $(E,J,\Omega,S,T)=$(0.776, 0.337, 1.38, 0.798, 0.213), (3.95, 2.67, 1.32, 0.987, 0.111), (16.9, 13.0, 1.19, 1.38, 0.0203), respectively.
We see that $\alpha(r)$ is deformed near the horizon and recovers the asymptotically AdS behavior as $r \to \infty$.
Comparing Figs.~\ref{geon_sol} and \ref{BR_sol}, we find that the function profiles outside the horizon look similar, especially in $r \gtrsim 0.8$ in this case.
This supports the notion that a black resonator is a dressed black hole where a black hole is placed inside a geon.
In the bottom-right panel of Fig.~\ref{BR_sol}, the norm of the Killing vector $g_{\mu\nu}K^\mu K^\nu$ is shown for the black resonators.
Near the horizon, it is negative and the black resonator is stationary near the horizon.
In a far region, the spacetime becomes dynamical. 
This property of the black resonator has been predicted in~\cite{Kunduri:2006qa}.

\subsection{Thermodynamics}

\begin{figure}
\begin{center}
\includegraphics[scale=0.7]{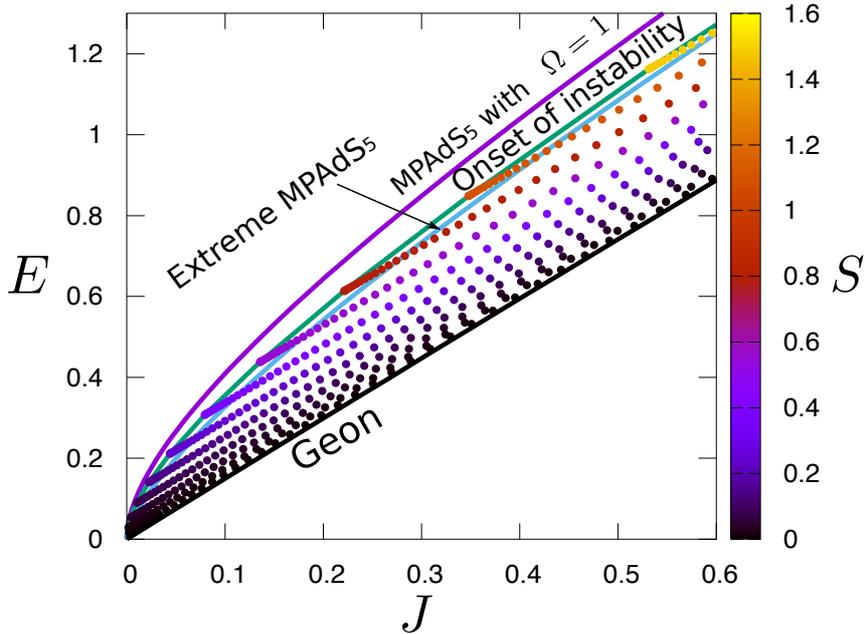}
\end{center}
\caption{%
Phase diagram of the MPAdS$_5$ black holes, black resonators, and geons.
The numerical data we constructed are marked with the dots in the $(E,J)$ plane.
The entropy $S$ is shown by the color map. 
The black, purple, green, and light blue curves correspond to the
geons, MPAdS$_5$ with $\Omega=1$,
MPAdS$_5$ at the onset of the superradiant instability, and 
extreme MPAdS$_5$, respectively.
}
\label{BR1}
\end{figure}

\begin{figure}
\begin{center}
\includegraphics[scale=0.5]{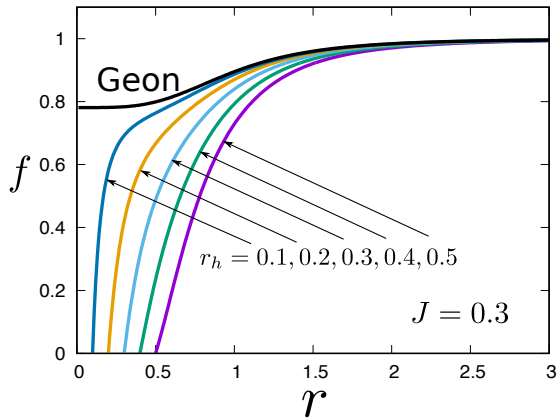}
\end{center}
\caption{%
The metric component $f(r)$ for $r_h=0.1,0.2,\cdots,0.5$.
The angular momentum is fixed to $J=0.3$.
}
\label{BRtoGeon}
\end{figure}

Fig.~\ref{BR1} shows the phase diagram of the MPAdS$_5$ black holes, black resonators, and geons on the $(E,J)$ plane.
The dots correspond to the data points that we numerically constructed.
The black resonators branch from the onset of the superradiant instability of the MPAdS$_5$ black holes and can be found even in the region where no regular MPAdS$_5$ solutions exist.
The black curve locating as the lower boundary is the family of the geons.
The black resonators interpolate the onset of the instability of the MPAdS$_5$ black holes and the geons.
The entropy of the black resonators is shown as a color map. It approaches zero near the geon limit.
In Fig.~\ref{BRtoGeon}, 
we compare the metric component $f(r)$ for the black resonators with different $r_h$ at a fixed $J=0.3$. 
This demonstrates that the functional profiles of the black resonators approach that of the geon as $r_h$ decreases.
For the numerical solutions, we monitored $\delta E-T\delta S - \Omega \delta J$ and found that 
this is consistent with zero within numerical accuracy.

\begin{figure}[t]
\centering
\includegraphics[scale=0.6]{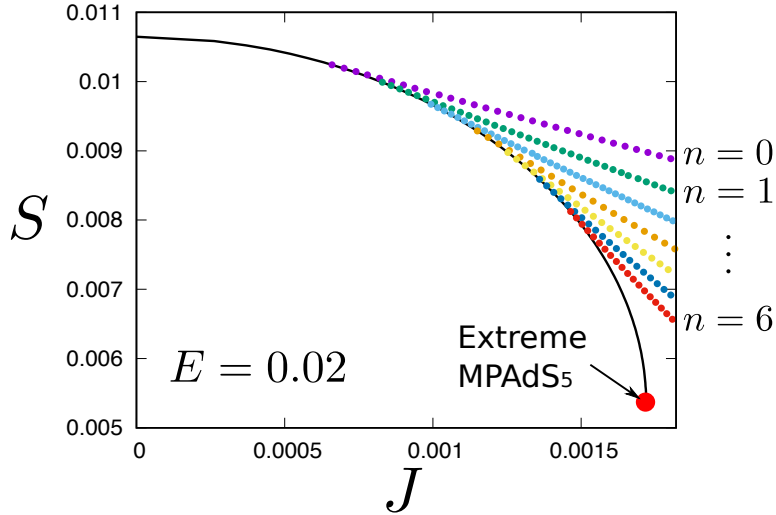}
\caption{Comparison of $S$ between the black resonators (dots) with $n=0,1,\cdots,6$ 
and MPAdS$_5$ (black curve) for the same $(E,J)$. The energy is fixed at $E=0.02$ and the angular momentum is varied.}
\label{fig:gvsmp_ES}
\end{figure}

The black resonators have larger entropy than the MPAdS$_5$ in the coexistence region as shown in Fig.~\ref{fig:gvsmp_ES}.
Families of black resonators branch from the MPAdS$_5$, and we plot 
data for the $n \le 6$ tones.
Extrapolated to arbitrary $n$, this result suggests infinite violation of uniqueness even for the black holes with the $R\times SU(2)$-symmetry.
We find that the entropies of the black resonators are always larger than that of the MPAdS$_5$.
Comparing the entropies of the black resonators between 
the data in the figure,
we find that the fundamental tone ($n=0$) has the largest entropy.
This result indicates that, in dynamical processes, 
the MPAdS$_5$ causing the superradiant instability can develop into the black resonators with $n=0$ 
if the $SU(2)$-symmetry is imposed.

\begin{figure}
  \centering
  \subfigure[Angular velocity $\Omega$]
 {\includegraphics[scale=0.45]{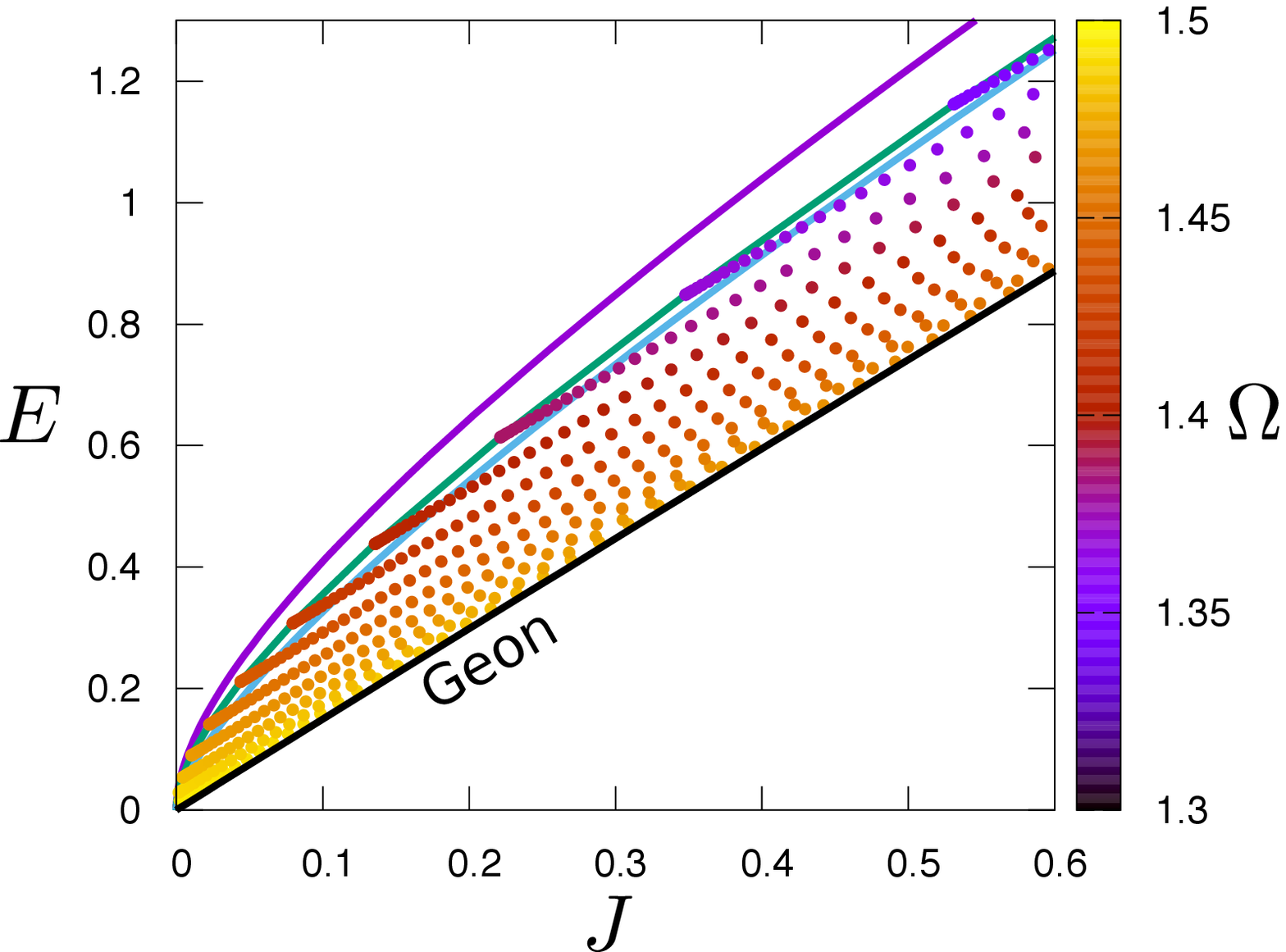}\label{EJO}
  }
\subfigure[Temperature $T$]
 {\includegraphics[scale=0.45]{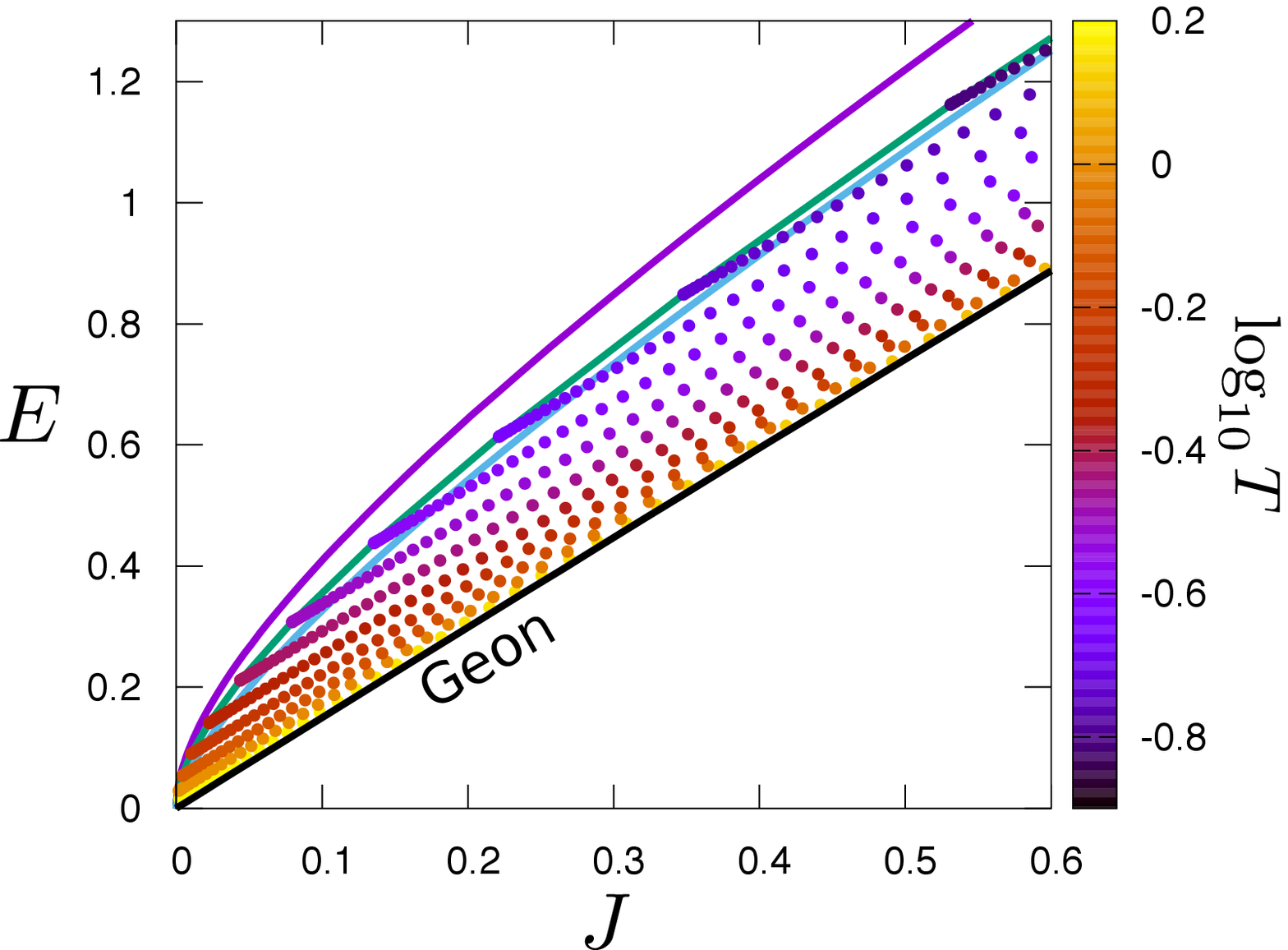}\label{EJT}
  }
\caption{%
The angular velocity $\Omega$ and temperature $T$ of the black resonators are shown as color maps on the $(E,J)$ plane.
\label{BR2} 
}
\end{figure}

Fig.~\ref{BR2} shows the angular velocity $\Omega$ and temperature $T$ of the black resonators as color maps.
The angular velocity is always greater than 1 and the horizon Killing vector becomes spacelike near the infinity $r = \infty$.
This indicates that the black resonators found in this paper would suffer from further superradiant instabilities, probably to 
$SU(2)$-violating modes~\cite{Green:2015kur,Niehoff:2015oga}.
In AdS$_4$, black resonators with $\Omega<1$ cannot exist because of the rigidity theorem~\cite{Hollands:2006rj,Moncrief:2008mr}.
In AdS$_5$, while the $U(1)$-symmetry generated by $\partial_\psi$ is broken, $\partial_\phi$ can be still preserved.
There may be a possibility of the existence of the black resonators with $\Omega<1$, but so long as we studied, we did not find any in AdS$_5$.
The temperature increases near the geons because the horizon radius is becoming small in the ``small black hole'' branch.

\begin{figure}[t]
\centering
\includegraphics[scale=0.6]{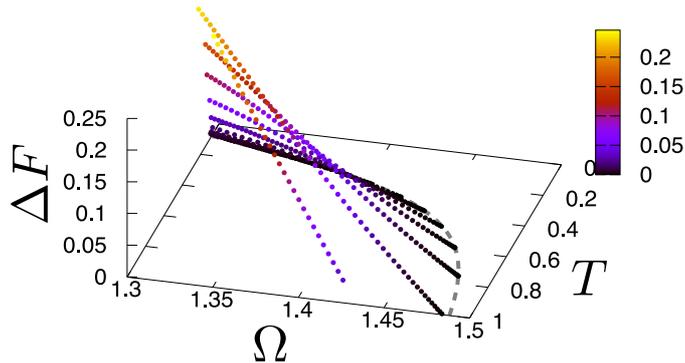}
\caption{The difference of the free energy of the $n=0$ black resonators from that of the MPAdS$_5$: $\Delta F \equiv F |_\mathrm{black \, resonator} - F |_\mathrm{MPAdS}$. We find $\Delta F > 0$ for $\alpha_0 \neq 1$. The onset of the superradiant instability, where $\Delta F=0$, is also shown in a gray dashed curve.}
\label{fig:gvsmp_WTFS}
\end{figure}

In Fig.~\ref{fig:gvsmp_WTFS}, we compare the free energy $F =E-TS-\Omega J$ 
between the MPAdS$_5$ and black resonators for the same $(T,\Omega)$.
Thermodynamically, the free energy of the black resonators is higher than that of the MPAdS$_5$ black holes.
Note that this comparison is in the small black hole region, and in any case both of these solutions have higher free energies than the thermal AdS, which would be the grand state.

\section{Conclusions}
\label{sec:con}

We constructed the black resonators emerging from the onset of 
a superradiant instability of the MPAdS$_5$ with equal angular momenta.
Geons were also obtained in their horizonless limit, which are nonlinear extensions of the normal modes of the global AdS.
We used a cohomogeneity-1 metric ansatz for the black resonators and geons, which has a $SU(2)$ symmetry.
It reduces the Einstein equations to ODEs and allows us to study the black resonators and geons 
in an extensive region in parameters.
We demonstrated that the black resonators and geons have a helical Killing vector and are time periodic.
We computed their thermodynamic quantities and 
obtained the phase diagram of the black resonators, geons, and MPAdS$_5$.
The black resonators exist even in the region where no regular MPAdS$_5$ solutions do.
They connect the onset of a superradiant instability of the MPAdS$_5$ and the geons.
They have larger entropies than the MPAdS$_5$ in the coexistence region. 
We also constructed the black resonators with overtones and found that  
they have lower entropies than the fundamental tone.
This indicates that, in dynamical processes, 
a MPAdS$_5$ can evolve into a black resonator with the fundamental tone if the $SU(2)$ symmetry is imposed.
As far as we studied, the angular velocity of the black resonator always satisfies $\Omega>1$.

What is the dual picture of the black resonator in the context of the AdS/CFT duality?
As shown in Eq.(\ref{Tmunu_non-rot}), in the non-rotating frame, the boundary stress tensor is time-periodic.
This means that the state dual to the black resonator breaks the time translation symmetry to a discrete one.
This sounds analogous to a time crystal.
However, there is a no-go theorem about realization of time crystals in the ground state~\cite{Watanabe:2014hea}.
We have found in Fig.~\ref{fig:gvsmp_WTFS} that the black resonators are not thermodynamically dominant.
This situation is consistent with the no-go theorem.

A benefit of this work's approach is the easiness to handle the dynamics corresponding to the superradiance because of the cohomogeneity-1 metric.
As shown in \cite{Green:2015kur}, 
the black resonators with $\Omega>1$ are unstable, and those we constructed will also develop into other spacetimes.\footnote{%
It was conjectured that their endpoint may not be described within classical gravity \cite{Niehoff:2015oga}. 
}
What perturbations will the black resonators be unstable to? 
How strong is the instability?
To answer such questions, it is necessary to work on stability analysis.
For the black resonators in AdS$_4$~\cite{Dias:2015rxy}, the perturbation equations will still be 3-dimensional PDEs.
For the cohomogeneity-1 black resonators found in this paper, these will reduce to ODEs, and
this will help us to carry out stability analysis directly.
If we find instabilities of the black resonators, 
these will suggest a new family of black hole solutions which have multi-frequencies~\cite{Choptuik:2018ptp}.

Investigating nonlinear time evolution of the superradiant instability would be an important future direction.
Dynamical processes involving superradiant instabilities have been recently considered 
in $(1+3)$-spacetime dimensions by numerical time evolution in \cite{Chesler:2018txn}.
If we impose the $SU(2)$-symmetry on dynamical spacetime as we have done in Eq.(\ref{metricanz}), 
we will just need to solve the time evolution of $(1+1)$-dimensional PDEs.
With such a simplification, we will be able to partially answer the question about the final fate of the superradiant instability.
In a similar way, we will be also able to study the weakly turbulent instability of AdS~\cite{Bizon:2011gg} 
in the pure gravity setup~(\ref{EHaction}).
The gravitational weakly turbulent instability without a rotation 
has been studied in~\cite{Bizon:2017yrh}.
(See also~\cite{Dias:2011ss} for a perturbative study.)
Recently, the generalization beyond preserving the spherical symmetry has also been considered \cite{Dias:2016ewl,Rostworowski:2017tcx,Dias:2017tjg,Bantilan:2017kok,Choptuik:2017cyd}.
It would be favorable to take into account the effect of rotations also from our perspective.

\acknowledgments
The authors would like to thank Akihiro Ishibashi, Masashi Kimura, Harvey Reall, Jorge Santos, Takahiro Tanaka, and Benson Way for useful discussions and comments.
We also thank the Yukawa Institute for Theoretical Physics at Kyoto University. 
Discussions during the YITP workshop YITP-T-18-05 ``Dynamics in Strong Gravity Universe'' were useful to complete this work.
The work of T.~I.~was supported in part by the Netherlands Organisation for Scientific Research (NWO) under the VIDI grant 680-47-518 and the Delta-Institute for Theoretical Physics ($\Delta$-ITP), which is funded by the Dutch Ministry of Education, Culture and Science (OCW), and in part by JSPS KAKENHI Grant Number 18H01214.
The work of K. M. was supported by JSPS KAKENHI Grant Number 15K17658 and 
in part by JSPS KAKENHI Grant No. JP17H06462.

\appendix

\section{Charged tensor and vector harmonics on $CP^1$}
\label{chargedTV}

In section~\ref{SRMPAdS}, we introduced the metric perturbation \eqref{da_pert} which has the $U(1)$-charges $\pm2$.
In this appendix, we consider that perturbation from the viewpoint of the $CP^1$ base.
In particular, we explicitly construct charged tensor harmonics on $CP^1$ and show that the perturbation with  
the $U(1)$-charges $\pm2$ corresponds to the lowest charged tensor modes.\footnote{%
The charged harmonics on $CP^1$ are also referred to as the monopole spherical harmonics.
The scalar and vector monopole spherical harmonics are constructed in~\cite{Wu:1976ge,Weinberg:1993sg}.
See also~\cite{Ishii:2008tm} for an explanation that the monopole spherical harmonics are related to the spherical harmonics on $S^3$.
Hence we expect that the perturbation \eqref{da_pert} has its counterpart in the viewpoint of the base space.
}

The Fubini-Study metric and K\"{a}hler potential on $CP^1$ are given by
\begin{equation}
 \hat{g}_{ij}dx^i dx^j=\frac{1}{4}(d\theta^2+\sin^2\theta d\phi^2)\ ,\quad
 A = \frac{1}{2}\cos\theta d\phi\ ,
\end{equation}
where $i,j=\theta,\phi$.
The K\"{a}hler form is defined as $J=dA$.
The conditions for a tensorial function $Y_{ij}$ on $CP^1$ to be charged tensor harmonics are
\begin{equation}
 D^i Y_{ij}=0\ ,\quad \hat{g}^{ij}Y_{ij}=0\ ,
\label{tcond}
\end{equation}
and
\begin{equation}
 (D^2+\lambda_T)Y_{ij}=0\ ,
\label{teigeneq}
\end{equation}
where $D_i=\hat{\nabla}_i-im A_i$ with $\hat{\nabla}$ being the covariant derivative with respect to $\hat{g}_{ij}$, 
and $\lambda_T$ denotes the eigenvalue of $-D^2=-D^i D_i$.
Eq.(\ref{tcond}) corresponds to the transverse and traceless (TT) condition. 
The charge $m$ corresponds to the eigenvalue of $-i\partial_{\chi/2}$ ($\chi/2 \in [0,2\pi)$)
and must be an integer: $m\in \bm{Z}$.\footnote{In the literature including \cite{Wu:1976ge}, the charge $q=m/2$ is also used which takes the values in $q \in \bm{Z}/2$.}

A naive argument for non-existence of charged tensor harmonics could be given as follows. 
The number of the components of $Y_{ij}$ is 3, and this
equals the number of the TT condition (\ref{tcond}).
Therefore, $Y_{ij}$ is completely determined just by the TT condition, and it would not 
further satisfy the eigenvalue equation~(\ref{teigeneq}).
However, we will show that it actually does, and we will also explicitly obtain the charged tensor harmonics on $CP^1$.

Let us introduce the ansatz for the tensor harmonics as
\begin{equation}
Y_{ij}=e^{in\phi}
\begin{pmatrix}
p(\theta) & q(\theta)\\
q(\theta) & -p(\theta)\sin^2\theta
\end{pmatrix}
\ .
\end{equation}
Here, $n$ is half-integer (integer) when $m$ is odd (even) because of the periodicity~(\ref{phipsi_id}).
This trivially satisfies the traceless condition in Eq.(\ref{tcond}). 
From the transverse condition, we obtain
\begin{equation}
\begin{split}
&p'(\theta)+2\cot \theta \, p(\theta) +i \, \frac{2n-m \cos\theta}{2\sin^2\theta}\, q(\theta)=0\ ,\\
&q'(\theta)+2\cot \theta \, q(\theta) -\frac{i}{2}\, (2n-m\cos\theta)\, p(\theta)=0\ .
\end{split}
\end{equation}
These can be analytically solved by
\begin{equation}
 \begin{pmatrix}
p(\theta)\\
q(\theta)
\end{pmatrix}
=
c_1
 \begin{pmatrix}
i\,\cos^{a-2}(\theta/2)\,\sin^{b-2}(\theta/2)\\
2\,\cos^{a-1}(\theta/2)\,\sin^{b-1}(\theta/2)
\end{pmatrix}
+c_2
 \begin{pmatrix}
-i\,\cos^{-a-2}(\theta/2)\,\sin^{-b-2}(\theta/2)\\
2\,\cos^{-a-1}(\theta/2)\,\sin^{-b-1}(\theta/2)
\end{pmatrix}
\ .
\label{Yt}
\end{equation}
where $a=m/2+n$ and $b=m/2-n$. From the regularity at $\theta=0$ and $\pi$, 
we find that $a$ and $b$ must be integers and satisfy 
\begin{equation}
 (a,b)\in \{a\geq 2\ ,b\geq 2\} \quad \textrm{or}\quad (a,b)\in \{a\leq -2\ ,b\leq -2\}\ .
\label{abcond}
\end{equation}
For the former and later cases, we need to set $c_2=0$ and $c_1=0$, respectively.
The condition Eq.(\ref{abcond}) can be equivalently rewritten as 
\begin{equation}
 |m|=2|n|+4+2k \quad (k=0,1,2,\cdots)\ .
\label{tmn}
\end{equation}
In this way, we can completely determine $Y_{ij}$ just from the TT conditions.

What is more, by a direct calculation, we can check that $Y_{ij}$ also ``accidentally'' 
satisfies the eigenvalue equation~(\ref{teigeneq}), and the eigenvalue is
\begin{equation}
 \lambda_T=2|m|-8\ .
\end{equation}
Using Eq.(\ref{tmn}), we find that
the charged tensor harmonics on $CP^1$ exists only for $|m|\geq 4$. 
This is consistent with the fact that there is no uncharged tensor harmonics on $CP^1$.
We can also explicitly check that
\begin{equation}
 J^{ij}D_i Y_{jk}=0\ .
\end{equation}
Therefore, all charged tensor harmonics on $CP^1$ are doubly transverse~\cite{Kunduri:2006qa}.
For $CP^{N\geq 2}$, there is an infinite infinite set of charged tensor harmonics for a fixed $m$~\cite{Kunduri:2006qa}.
Meanwhile, for $CP^1$, there are only finite numbers of the harmonics satisfying Eq.(\ref{tmn}).

The gravitational perturbation~(\ref{da_pert}) is then identified as a charged tensor perturbation on $CP^1$.
We can write $\sigma_-^2$ in the components on $S^2$ as
\begin{equation}
 (\sigma_-)_i  (\sigma_-)_j=\frac{i}{4}e^{2i\chi} 
\begin{pmatrix}
i & \sin\theta\\
\sin\theta & -i\sin^2\theta 
\end{pmatrix}
\ ,
\end{equation}
where the factor $e^{2i\chi}$ is for the $S^1$ fiber.
This corresponds to the charged tensor harmonics with $m=4$ and $n=0$.
Similarly, $\sigma_+^2$ corresponds to $m=-4$ and $n=0$.

In a similar way, we can also explicitly construct the doubly transverse charged vector harmonics on $CP^1$.
(Those on $CP^{N\geq 2}$ have been studied in \cite{Durkee:2010ea}.)
As the solution of the doubly transverse conditions for a vectorial function $Y_i$, $D^i Y_i=J^{ij} D_i Y_j=0$, we obtain
\begin{equation}
Y_i=
c_1
\begin{pmatrix}
i\cos^{a-1}(\theta/2)\sin^{b-1}(\theta/2) \\
2\cos^{a}(\theta/2)\sin^{b}(\theta/2)
\end{pmatrix}
+
c_2
\begin{pmatrix}
-i\cos^{-a-1}(\theta/2)\sin^{-b-1}(\theta/2) \\
2\cos^{-a}(\theta/2)\sin^{-b}(\theta/2)
\end{pmatrix}
\ .
\end{equation}
From the regularity at $\theta=0$ and $\pi$, we have
\begin{equation}
 |m|=2|n|+2+2k \quad (k=0,1,2,\cdots) \ .
\label{vmn}
\end{equation}
This satisfies the eigenvalue equation $(D^2+\lambda_V)Y_i=0$ with 
\begin{equation}
 \lambda_V=2|m|-4\ .
\end{equation}
They exist only for $|m|\geq 2$.

\section{Perturbative construction of the geons}
\label{pGeon}

In this appendix, we explain the construction of the perturbative geon results \eqref{EJOmega_pert}.

We consider a perturbative expansion of 
$\Phi(r)=(f(r),g(r),h(r),\alpha(r),\beta(r))^T$ as
\begin{equation}
 \Phi(r)=\sum_{m=0}^\infty \Phi^{(m)}(r)\epsilon^{m}\ ,
\end{equation}
where $\epsilon$ is a small parameter. 
As the background, we take the global AdS$_5$ in the rotating frame: 
$\Phi^{(0)}=(1,1,\Omega^{(0)},1,1)^T$. 
In section~\ref{SRMPAdS}, we showed that $\alpha^{(1)}=\delta \alpha$ satisfies 
the decoupled equation~(\ref{da_eq}). 
It can be also easily checked that $f^{(1)}=g^{(1)}=h^{(1)}=\beta^{(1)}=0$ in the leading order.
The exact solution for $\alpha^{(1)}$ is given by Eq.(\ref{da_exact}).
We focus on the fundamental tone: $\Omega^{(0)}=3/2$.
Then, we obtain
\begin{equation}
 \alpha^{(1)}(r)=\frac{r^2}{(1+r^2)^3}\ .
\end{equation}

We then go to the next order.
The second order equations are written in the form
\begin{equation}
 L\Phi^{(2)}= S^{(2)}\ ,
\label{2ndeq}
\end{equation}
where
\begin{equation}
L=
 \begin{pmatrix}
L_{11}&L_{12}&0&0&L_{15}  \\
0&L_{22}&0&0&L_{25}  \\
0&0&L_{33}&0&0  \\
0&0&0&L_{44}&0  \\
0&0&0&0&L_{55}  
\end{pmatrix}
\end{equation}
with
\begin{equation}
\begin{split}
&L_{11}=\partial_r\ ,\quad
L_{12}=\frac{2(1+2r^2)}{r(1+r^2)}\ ,\quad 
L_{15}=\frac{2+3r^2}{3(1+r^2)}\partial_r + \frac{2}{3r(1+r^2)}\ ,\\
&L_{22}=\partial_r+\frac{2(1+2r^2)}{r(1+r^2)}\ ,\quad
L_{25}=\frac{1}{3(1+r^2)}\partial_r + \frac{10}{3r(1+r^2)}\ ,\\
&L_{33}=\partial_r^2+\frac{5}{r}\partial_r\ ,\quad
L_{44}=\partial_r^2+\frac{3+5r^2}{r(1+r^2)}\partial_r - \frac{4(2-7r^2)}{r^2(1+r^2)^2}\ ,\\
&L_{55}=\partial_r^2+\frac{3+5r^2}{r(1+r^2)}\partial_r - \frac{8}{r^2(1+r^2)}\ .
\end{split}
\end{equation}
The source term $S^{(2)}$ is given by
\begin{multline}
 S^{(2)}=\bigg(
-\frac{2r^3(1-4r^2)}{(1+r^2)^7},
\frac{2r^3(7-4r^2)}{(1+r^2)^7},
\frac{24r^2}{(1+r^2)^7},\\
\frac{2r^2(4-15r^2+8r^4)}{(1+r^2)^8},
-\frac{20r^2}{(1+r^2)^7}
\bigg)^T\ .
\end{multline}
The second order equation~(\ref{2ndeq}) is solved by
\begin{multline}
\Phi^{(2)}=\bigg(
-\frac{3+18 r^2+40 r^4+20 r^6+4 r^8}{45(1+r^2)^6}, 
-\frac{r^2(30-25r^2+22r^4+5r^6)}{90(1+r^2)^6},\\
\Omega^{(2)}-\frac{2+10r^2+5r^4+r^6}{20(1+r^2)^5},
\frac{r^4}{2(1+r^2)^6},
\frac{r^2(10+5r^2+r^4)}{30(1+r^2)^5}
\bigg)^T\ ,
\label{Phi2}
\end{multline}
where $\Omega^{(2)}$ is an integration constant corresponding to 
the second order deviation of the angular velocity from the global AdS$_5$.
This is actually determined from the third order analysis as we will see shortly. 
Other integration constants are determined by imposing the regularity at $r=0$ and asymptotically AdS conditions~(\ref{AsympAdS}).

The third order equation is given by
\begin{equation}
 L\Phi^{(3)}= S^{(3)}\ ,
\end{equation}
where the $\alpha$-component of $S^{(3)}$ is
\begin{equation}
 S^{(3)}_\alpha = \frac{48\Omega^{(2)} r^2}{(1+r^2)^5} +\frac{2r^2(6-44r^2+637r^4+42r^6-17r^8-20r^{10})}{45(1+r^2)^{11}}\ ,
\end{equation}
and the other components are zero.
Therefore, $f^{(3)}=g^{(3)}=h^{(3)}=\beta^{(3)}=0$ are solutions to the third order.
The equation of $\alpha^{(3)}$ can be analytically solved.
From the asymptotically AdS condition $\alpha^{(3)}\to 0$ $(r\to \infty)$, the second order deviation of the angular velocity is determined as
\begin{equation}
 \Omega^{(2)}=-\frac{1}{180}\ .
\label{Omega2nd}
\end{equation}
From Eq.(\ref{Phi2}), the asymptotic forms of the second order perturbations are given by
\begin{equation}
 f^{(2)}=-\frac{4}{45r^4}+\cdots\ ,\quad
 h^{(2)}=\Omega^{(2)}-\frac{1}{20r^4}+\cdots\ ,\quad
 \beta^{(2)}=\frac{1}{30r^4}+\cdots\ .
\end{equation}
The second order deviations of $c_f$, $c_h$ and $c_\beta$, defined in Eq.(\ref{asym}), are hence
$c_f^{(2)}=-4/45$, $c_h^{(2)}=-1/20$ and $c_\beta^{(2)}=1/30$.
Then, from Eq.(\ref{EJdef}), 
the second order contributions for the mass and angular momentum are obtained as
\begin{equation}
 E^{(2)}=\frac{\pi}{8}\times \frac{3}{10}\ ,\quad J^{(2)}=\frac{\pi}{8}\times \frac{1}{5}\ .
\end{equation}

We can go to the higher orders by continuing the same procedure. 
As the fourth order equation, 
we have $L\Phi^{(4)}=S^{(4)}$ where all components of $S^{(4)}$ are non-zero.
We can solve the equation analytically and find that $\Omega^{(4)}$, which is a constant term in $h^{(4)}$, is not determined from the boundary conditions in this order.
However, at the fifth order, we obtain a decoupled equation for $\alpha^{(5)}$. Imposing $\alpha^{(5)}\to 0$ $(r\to \infty)$, we obtain
\begin{equation}
 \Omega^{(4)}=-\frac{356}{2338875}\ .
\label{Omega4th}
\end{equation}
From the asymptotic behaviour of the fourth-order perturbative solution with the above $\Omega^{(4)}$, we obtain the fourth-order deviation of the mass and angular momentum as
\begin{equation}
E^{(4)}=\frac{\pi}{8}\times \frac{803}{84000}\ ,\quad
J^{(4)}=\frac{\pi}{8}\times \frac{2549}{378000}\ .
\end{equation}
Putting together the above results, we obtain Eq.(\ref{EJOmega_pert}).

\bibliography{bunken_BR}

\end{document}